\documentclass[11pt]{IEEEtran2e} %
\pdfoutput=1
\usepackage{times}
\usepackage{amsmath}
\usepackage{subfigure}
\usepackage{colordvi}
\usepackage{color}
\usepackage{subfigure}
\usepackage{epsfig}
\usepackage{amsfonts}
\usepackage{graphicx}
\usepackage{latexsym}
\usepackage{delarray}

\newcommand{\urlBiBTeX}[1]{\url{#1}}
\newcommand{\urlbibteX}[1]{\url{#1}}
\def\BibTeX{{\rm B\kern-.05em{\sc i\kern-.025em b}\kern-.08em
    T\kern-.1667em\lower.7ex\hbox{E}\kern-.125emX}}

\advance \hoffset 0.3in 
\advance \voffset 0.6in

\newtheorem{theorem}{Theorem}

\newtheorem{lemma}{Lemma}

\def\Ain{A}
\def\Aout{D}

\def\conv{*}

\def\eps{\varepsilon}
\def\E{{E}}  %
\def\G{{\cal G}}

\def\P{{Pr}} %
\def\S{{\cal S}}

\def\htss{\textit{htss }}
\def\htt{\textit{ht }}

\begin{document}
\date{}

\title{Delay Bounds for Networks with Heavy-Tailed and Self-Similar Traffic}

\author{
J\"{o}rg Liebeherr, ~Almut Burchard,
~Florin Ciucu
\vspace{-4mm}
\thanks{
J. Liebeherr ({\tt jorg@comm.utoronto.ca}) is with the Department of Electrical and Computer Engineering, University of Toronto. A. Burchard ({\tt almut@math.utoronto.ca}) is with the Department of Mathematics, University of Toronto.
F. Ciucu ({\tt florin@net.t-labs.tu-berlin.de}) is with Deutsche Telekom Laboratories at TU Berlin.
}
\thanks{
The research in this paper is supported in part by the National
Science Foundation under grant CNS-0435061,  by two NSERC Discovery
grants and an NSERC Strategic grant.}
}

\maketitle
\begin{abstract}
We provide upper bounds on the end-to-end backlog and delay 
in a network with  heavy-tailed and self-similar traffic. 
The analysis follows a network calculus approach 
where traffic is characterized by envelope 
functions  and service is described by  
service curves. A key contribution of this paper is 
the derivation of a probabilistic sample path bound 
for heavy-tailed self-similar arrival processes, which is enabled 
by a suitable envelope characterization, 
referred to as \htss envelope. 
We derive a heavy-tailed service curve for 
an entire network path when the service at each node on the path is 
characterized by heavy-tailed service curves. We obtain  
backlog and delay bounds for traffic that is 
characterized by an \htss envelope and receives service given 
by a heavy-tailed service curve. 
The derived performance bounds are non-asymptotic in that they 
do not assume a steady-state, large buffer, or many sources regime.
We also explore the scale of growth of delays as a function of the 
length of the path. 
The appendix contains an analysis for self-similar traffic with a 
Gaussian  tail distribution.
\end{abstract}

\section{Introduction} 

Traffic measurements in the 1990s  provided evidence of 
self-similarity  in aggregate network traffic  
\cite{LeWi94}, and heavy-tailed files sizes and bursts 
were found to be among the root causes  \cite{CrovellaB97,Willinger95}.
Since such traffic 
induces backlog and delay distributions 
whose tails decay slower than exponential,
the applicability of analytical techniques based on Poisson or 
Markovian traffic models in network 
engineering  has been called into question \cite{Paxson95}, thus 
creating a need for new approaches to teletraffic theory. 

A random process $X$ is said to have a {\em heavy-tailed} distribution 
if its tail distribution is governed by a power-law  
$\P(X (t) >x)\sim Kx^{-\alpha}$,
with a tail index $\alpha \in (0, 2)$ and a scaling constant $K$.\footnote{We 
write $f(x)\sim g(x)$,  if $\lim_{x\rightarrow\infty}f(x)/g(x)=1$.}
We will consider tail indices in the range $1<\alpha<2$, 
where the distribution has a finite mean, but infinite variance. 
A random process $X$  is said to be {\em self-similar} if 
a properly rescaled version of the process has the 
same distribution as the original process. We can write this 
as $X(t) \sim_{dist} a^{-H} X(at)$ for every $a>0$. 
The exponent $H \in (0,1)$, referred to as the {\em Hurst parameter}, specifies 
the degree of self-similarity.\footnote{The networking literature frequently 
uses the weaker concept of {\em second-order self-similarity}. 
Since we will work with heavy-tailed distributions, for which higher moments are not 
available, we use the more general definition of self-similarity.}
We refer to a process as heavy-tailed self-similar if it satisfies both 
criteria. 

A performance analysis of networks with heavy-tailed self-similar 
traffic or service, where no higher moments are available, 
is notoriously hard, especially an analysis of a 
network path across multiple nodes. 
Single node queueing systems with 
heavy-tailed processes have been studied extensively 
\cite{BoxmaD98,Liu05,Park2000}. 
However, there exist only few works that can be 
applied to analyze multi-node paths. 
These works generally consider an asymptotic regime 
with large buffers, many sources, or in the steady state. 
Tail asymptotics for multi-node networks have been derived for various 
topologies, such as feedforward networks \cite{Huang99}, 
cyclic networks \cite{Ayhan04}, 
tandem networks with identical service times \cite{Boxma79}, 
and tandem networks where packets 
have independent service times at nodes 
in the more general context of stochastic event graphs \cite{BaccelliLF04}. 
The accuracy of some asymptotic approximations has been called into question, 
particularly, the quality of large buffer asymptotics for heavy-tailed  
service distributions was found to be lacking in \cite{Abate94}, thus 
motivating a performance analysis in a non-asymptotic regime. 

This paper presents a non-asymptotic delay 
analysis for multi-node networks  with heavy-tailed self-similar traffic 
and heavy-tailed service. 
We derive the bounds for a flow or flow aggregate that traverses a network path and  
experiences cross traffic from heavy-tailed self-similar traffic at each node. 
Both fluid and packetized interpretations of service are supported; in 
the latter case, we assume that a packet maintains the same size at each traversed node.
A key contribution of this paper is 
a probabilistic sample path bound 
for heavy-tailed self-similar arrival processes. 
The derivation of the sample path bound is made possible by a suitable 
envelope characterization for heavy-tailed self-similar traffic, 
referred to as \htss envelope. 
We present a characterization for heavy-tailed service and show 
that it can express end-to-end service available on a path 
as a composition of the heavy-tailed service at each node. 
Our end-to-end service 
characterization enables the computation of 
end-to-end delay  bounds using our single-node result. 
In an asymptotic regime, our bounds  follow (up to a logarithmic correction) the same  power
law tail decay as asymptotic results that exist in the literature for 
single nodes. 
Finally, we show that end-to-end delays of heavy-tailed traffic and service 
grow polynomially with the number of nodes. 
For example, for a Pareto traffic source with tail index $\alpha$ 
we find that end-to-end delays are bounded by 
${O}(N^{\frac{\alpha+1}{\alpha-1}}  (\log N)^\frac{1}{\alpha-1})$
in the number of nodes $N$. 

Our analysis follows a network calculus 
approach where traffic is characterized in terms of {\em envelope functions}, 
which specify upper bounds on traffic over time intervals, 
and service is characterized by {\em service curves}, 
which provide lower bounds on the service available to 
a flow \cite{Book-Leboudec}.  
An attractive feature of the network calculus is that the service 
available on a path can be 
composed from service characterizations for each node of the path. 
We consider a probabilistic setting that permits  
performance metrics to be violated with a small probability.  
Probabilistic extensions of the network calculus are available for 
traffic with exponential tail distributions \cite{Chang94}, 
distributions that decay faster than any polynomial \cite{StaSi00}, 
and traffic distributions with an effective bandwidth \cite{Chang94}. 
The latter two groups include certain self-similar processes, in particular, 
those governed by fractional Brownian motion \cite{Nor95},  
but do not extend to heavy-tailed distributions. 
There are also efforts for extending the network calculus to 
heavy-tailed  distributions \cite{Makra2000,Barbosa05,Jiang05a,Hatzinakos01}, 
which are discussed in more detail in the next section.

The remainder of this paper is organized as follows. In 
Section~\ref{sec:envelope} and Section~\ref{sec:service}, 
respectively, we discuss our characterization of 
heavy-tailed traffic and service by 
appropriate probabilistic bounds. 
In Section~\ref{sec:convolution} we present 
our main results: (1) a sample path envelope for 
heavy-tailed self-similar traffic, 
(2) probabilistic bounds for delay and backlog, 
(3) a description of the leftover capacity at a constant-rate link 
with heavy-tailed self-similar cross traffic, 
and (4) a composition result for service descriptions at multiple nodes. 
In Section~\ref{sec:scaling} we discuss the 
scaling properties of the derived delay bounds 
in terms of power laws.  
We present brief conclusions in Section~\ref{sec:conclusion}.

\section{The {\large\it htss} Traffic Envelope }\label{sec:envelope}

In this section we present and evaluate a probabilistic envelope function, 
for characterizing  heavy-tailed self-similar 
network traffic that permits the derivation of rigorous backlog and delay bounds.
The proposed \htss envelope further develops concepts that were previously studied 
in \cite{Barbosa05,Jiang05a,Hatzinakos01}.

We consider arrivals and departure of traffic at a system, which represents a 
single node or a sequence of multiple nodes. 
We use a continuous time model where arrivals and departures of a traffic flow 
at the system for a time interval $[0,t)$  are represented by
left-continuous processes $\Ain(t)$ and $\Aout(t)$,
respectively. The arrivals in the time interval $[s,t)$
are denoted by a bivariate process $\Ain(s,t):=\Ain(t)-\Ain(s)$.
Backlog and delay at a node are represented by 
$B(t)=\Ain(t)-\Aout(t)$ and
$W(t)=\inf\left\{d:\Ain(t-d)\leq\Aout(t)\right\}$, respectively.
When $\Ain$ and $\Aout$ are plotted as functions of time, 
$B$ and $W$ are the vertical and horizontal distance, respectively, 
between these functions. 

A {\em statistical envelope} $\G$ for an arrival process $\Ain$ is 
a non-random function which bounds arrivals over a 
time interval  
such that, for all $s, t \geq 0$ and for all $\sigma > 0$ \cite{CiBuLi06}: 
\begin{align}
\P \Big(A(s,t) > \G(t - s; \sigma)\Big)  \leq  \eps (\sigma) \ ,
\label{eq:localenv1}
\end{align}
where $\eps$ is a non-increasing function of $\sigma$ that satisfies
$\eps(\sigma)\to 0$ as $\sigma\to\infty$. The function $\eps(\sigma)$ is used as a bound 
on the violation probability. 
Statistical envelopes have been  developed for  many different 
traffic types, including regulated, 
Markov modulated On-Off, and Gaussian self-similar traffic. 
A recent survey provides an 
overview of envelope concepts  \cite{Mao06}.

The computation of performance bounds, e.g., bounds on backlog 
delay, and output burstiness, requires a statistical envelope that bounds 
an entire sample path $\left\{\Ain(s,t)\right\}_{s\leq t}$.
A {\em statistical sample path envelope} $\overline{\G}$ is a statistical 
envelope  that satisfies for all $t \geq 0$ and for all $\sigma > 0$ \cite{CiBuLi06}:
\begin{align}
\P \Big(\sup_{s \leq t } \Big\{A(s,t) - \overline{\G}(t - s; \sigma)\Big\}  > 0 \Big) 
\leq  \overline{\eps} (\sigma) \ .
\label{eq:env1}
\end{align}
Clearly, a statistical sample path envelope is also a statistical envelope, but not vice versa. In fact, only few statistical envelopes (in the sense of Eq.~(\ref{eq:localenv1})) lend themselves easily to the development of 
sample path envelopes  (as in  Eq.~(\ref{eq:env1})). 
One of the earliest such envelopes appears in 
the {\em Exponentially Bounded Burstiness} (EBB) model 
from \cite{Yaron93}, which 
requires that 
\(
\P(\Ain(s,t)>r(t-s)+\sigma)\leq M  e^{- a\sigma }, 
\)
for some constants $M$, $r$ and $a$ and 
for all $\sigma > 0$. If $r$ corresponds to the mean rate of 
traffic, an EBB envelope specifies 
that the deviation of the traffic flow from its mean rate has an exponential decay. 
A sample path bound for EBB envelopes in 
the sense of Eq.~(\ref{eq:env1}) is obtained via the union bound\footnote{For two events 
$X$ and $Y$, $\P(X \cup Y) \leq \P(X) + \P(Y)$.}
by evaluating  
the right-hand side of Eq.~(\ref{eq:env1}) as 
$\sum_{k} \P(\Ain(s_k,t)>\overline{\G}(t-s_{k-1};\sigma))$ for a suitable discretization $\{s_k\}_{k=1,2,\ldots}$, 
yielding $\overline{\G}(t-s;\sigma)= Rt + \sigma$ for $R>r$ and 
$\overline{\eps} (\sigma) = \frac{M e^{-a \sigma}}{1 - e^{-a (R-r)}}$ \cite{CiBuLi06}.
The EBB envelope has been generalized to distributions with moments of all orders, 
referred to as {\em Stochastically Bounded Burstiness 
(SBB)} \cite{StaSi00} and corresponding sample path bounds have been developed in \cite{Yin02}. 
SBB envelopes can characterize  
arrival processes that are self-similar, but not heavy-tailed. 
For instance, fractional Brownian motion processes can be fitted with an 
envelope function $\G (t) = r t + \sigma$
with a  Weibullian bound on the violation probability  
of the form $\eps(\sigma)=K e^{-(\sigma/a)^{\alpha}}$ for some $a>0$ and $0<b<1$.

A statistical envelope for general self-similar arrival processes 
can be expressed  as 
\begin{align}
\P \Big(A(s,t) > r(t-s)+\sigma (t-s)^H\Big)  \leq  \eps (\sigma) \ .
\label{eq:ss-env}
\end{align}
Note that $A(s,t) - r(t-s) \leq_{dist} X(t-s)$, where $X$ is a process 
satisfying the self-similar property given in the introduction. 
For self-similar traffic, it is natural to allow a heavy-tailed violation probability, 
since self-similarity can arise from  heavy-tailed arrival 
processes with independent increments.\footnote{This is evident in Eqs.~(\ref{eq:gclt1})--(\ref{eq:envPareto1}) below in an application of the Generalized Central Limit Theorem for a  Pareto traffic source.}
This consideration leads to our
proposed  extension of the EBB and SBB concepts 
that capture characteristics of heavy-tailed and self-similar
traffic. We define a {\em heavy-tail self-similar  (htss) 
envelope} as a bound that satisfies for all $\sigma > 0$ that 
\begin{equation}
\P\Big(\Ain(s,t)>r(t-s)+\sigma(t-s)^{H}\Big)\leq
K\sigma^{-\alpha}~.\label{eq:envelope}
\end{equation}  
where $K$ and $r$ are constants, and $H$ and $\alpha$, respectively, 
indicate the Hurst parameter  and the tail index. 
We generally assume that $\alpha\in(1,2)$, that is, arrivals have a finite 
mean but infinite variance, and $H\in\left(0,1\right)$. 
In the \htss envelope  the 
probability of deviating from the average rate $r$ follows a power law.
Moreover, due to self-similarity, these deviations 
may increase as a function of time. 
Since the \htss envelope specifies a bound, it can be used to describe any type of 
traffic, but the  characterizations will be loose unless the traffic 
has some heavy-tailed self-similar properties. 
In terms of Eq.~(\ref{eq:localenv1}), the \htss envelope is a statistical 
envelope with
\begin{equation}
\G(t; \sigma) = rt+\sigma t^{H} \ , ~~ 
\eps (\sigma) = K\sigma^{-\alpha}
\label{eq:htts}
\end{equation}
In Section~\ref{sec:convolution}, we will derive 
a sample path envelope for the \htss envelope, which is necessary for the  computation of 
probabilistic upper bounds on backlog and delay of heavy-tailed self-similar 
traffic at a network node. 

Characterizations of  self-similar and heavy-tailed traffic 
by envelopes have been presented before, generally, by exploiting 
specific properties of $\alpha$-stable processes  \cite{Makra2000,Hatzinakos01}.
An envelope for $\alpha$-stable processes in 
\cite{Barbosa05} takes the same form 
$\G(t; \sigma) = rt+\sigma t^{H}$ as the  \htss envelope, but specifies 
a fixed violation probability rather than a bound on the distribution.
An issue with such a characterization 
is that it does not easily lead to sample path 
envelopes.  
For $H=0$, a sample path version of Eq.~(\ref{eq:envelope}) has been 
obtained in \cite{Jiang05a} by applying an a-priori bound on the 
backlog process of an $\alpha$-stable self-similar process 
from \cite{Hatzinakos01}.
Since the backlog bound given in  Eq.~(24) of \cite{Hatzinakos01} 
is a lower bound (and not an upper bound) on the tail distribution 
of the buffer occupancy, the envelope in \cite{Jiang05a} does not satisfy 
Eq.~(\ref{eq:env1}). 
In Section~\ref{sec:convolution} it will become evident that sample path 
envelopes for arrivals and backlog bounds are interchangeable, in that 
the availability of one can be used to derive the other. Thus, 
the sample path bound derived in this paper 
for heavy-tailed self-similar processes  satisfying the 
\htss envelope from Eq.~(\ref{eq:envelope}) also   
provides the first rigorous backlog bounds for this general class of processes. 

\bigskip
In the remainder of this section, we show how to construct \htss envelopes 
for relevant distributions, as well as for measurements of packet traces. 
Ever since traffic measurements 
at Bellcore from the late 1980s discovered long-range dependence and self-similarity
in aggregate network traffic  \cite{LeWi94}, many studies have 
supported, refined, sometimes also repudiated (e.g., \cite{Gong05}) these findings. 
This report does not participate in the debate whether aggregate network traffic is 
best characterized as  short-range or long-range dependent, self-similar or  multi-fractal, 
short-tailed or heavy-tailed, and so on. 
Rather we wish to provide tools for evaluating the performance  of networks that 
may see heavy-tailed self-similar traffic, and 
shed light on the opportunities and pitfalls of envelope descriptions 
for heavy-tailed traffic. 

\subsection{$\alpha$-stable Distribution}
\label{sec:alphaStable} 

Stable distributions provide well-established models 
for non-Gaussian processes with infinite variance.  
The potential  of applying stable processes to data networking 
was demonstrated in  \cite{Hatzinakos01} by fitting 
traces of aggregate traffic (i.e., the Bellcore traces studied in \cite{LeWi94}) to 
an $\alpha$-stable self-similar process. 

A defining property of an $\alpha$-stable distribution ($0 < \alpha \leq 2$) 
is that the linear superposition of i.i.d. $\alpha$-stable random 
variables preserves the original distribution. 
That is, if $X_1,$ $X_2,$ $\ldots X_m$ are independent random variables 
with the same (centered) $\alpha$-stable distribution, 
then $m^{-1/\alpha} \sum_{i=1}^{m} X_i$ has the same distribution. 
A challenge of working with $\alpha$-stable distributions is 
that closed-form expressions for the distribution are only available for 
a few special cases. However, there exists an explicit expression for the characteristic function of 
stable distributions, in terms of four parameters (see \cite{Samo94}): 
a \textit{tail index} $\alpha\in(0,2]$, 
a \textit{skewness} parameter $\beta\in[-1,1]$, 
a \textit{scale} parameter $a>0$, and 
a \textit{location} parameter $\mu\in\mathbb{R}$. 
For our purposes it is sufficient to work with a normalized 
stable random variable $S_\alpha$ where $\beta =1$, $a = 1$, and $\mu = 0$. 

The point of departure for our characterization of $\alpha$-stable processes with \htss envelopes is the  $\alpha$-stable process proposed in \cite{Hatzinakos01} which takes the form 
\begin{equation}
\Ain(t)\stackrel{dist.}{=}rt+bt^{H}S_{\alpha}~.\label{eq:asss}
\end{equation}
Here, $r$ is the mean arrival rate and 
$b$ is a parameter that describes the dispersion around the mean.

\medskip	
\noindent
{\em Remark: }
We can use Eq.~(\ref{eq:asss}) to observe the statistical multiplexing gain 
of $\alpha$-stable processes. 
By the defining property of $S_\alpha$, the superposition of $N$ i.i.d. processes as 
in Eq.~(\ref{eq:asss}), denoted by  $\Ain_{\mbox{\small \sl mux}}$, yields 
\[
\Ain_{\mbox{\small \sl mux}}(t)=Nrt+ N^{1/\alpha} bt^{H}S_{\alpha}~.
\] 
Since $1/\alpha < 1$ in the considered range $\alpha \in (1,2)$, 
the aggregate of a set of flows increases 
slower than linearly in the number of flows, thus, 
giving clear evidence of multiplexing gain.  
The multiplexing gain diminishes as $\alpha \rightarrow 1$. 

\begin{figure}
\centerline{\includegraphics[width=0.4\textwidth]{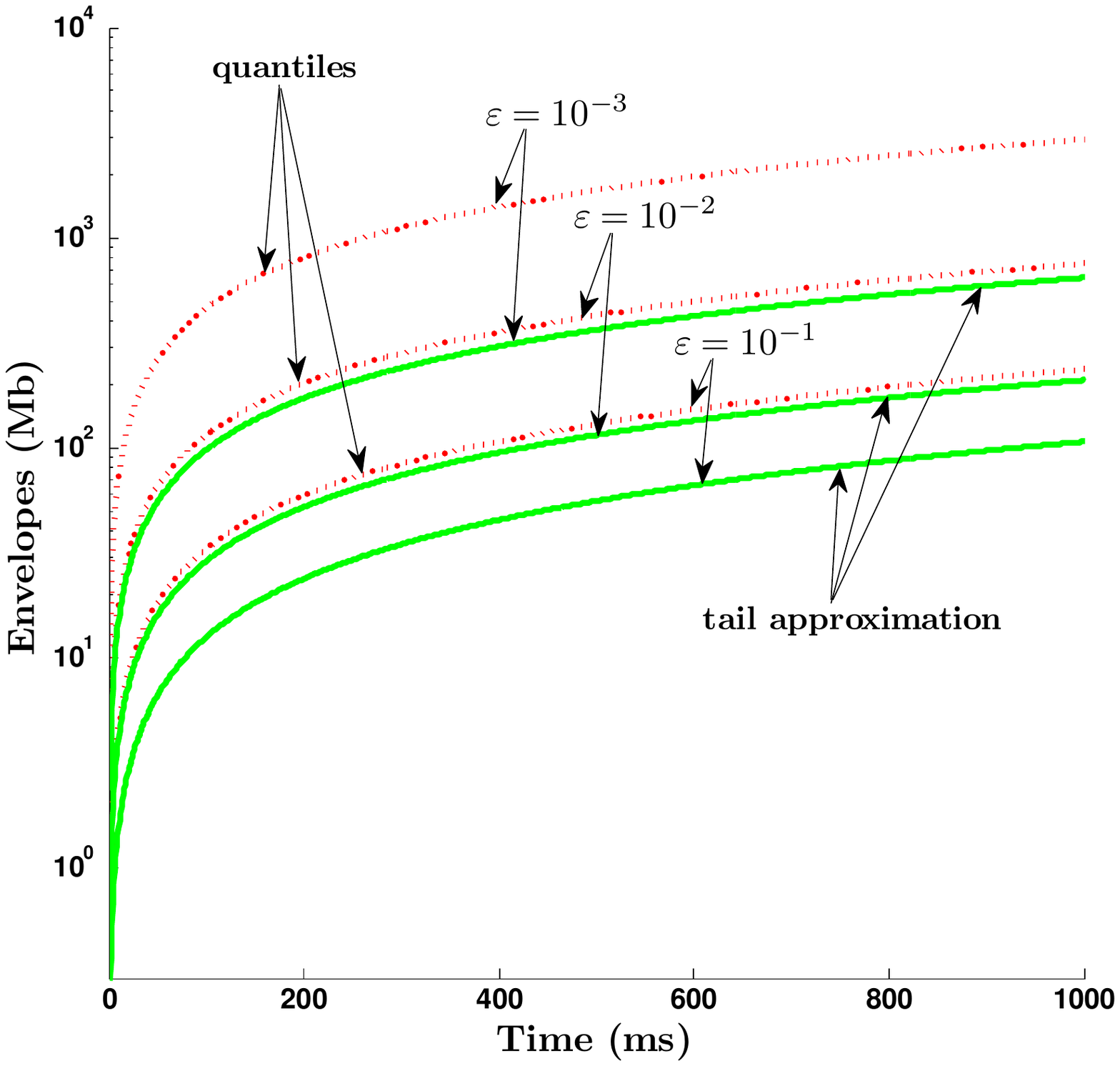}}
\caption{Comparison of \htss envelopes for an $\alpha$-stable distribution
with $r= 75~Mbps ,\alpha= 1.6,H= 0.8,b =  60~Mbps$.} \label{fig:alpha}
\end{figure}

We can obtain an \htss envelope for Eq.~(\ref{eq:asss}) from the tail
approximation for $\alpha$-stable distributions \cite{Nolan09} 
\begin{equation}
\P\Big(S_{\alpha}>\sigma\Big)\sim
\left(c_{\alpha}\sigma\right)^{-\alpha}~~,~\sigma\rightarrow\infty~,\label{eq:approxTailStable}
\end{equation}
where
$c_{\alpha}=\left(\frac{2\Gamma(\alpha)\sin{\frac{\pi\alpha}{2}}}{\pi}\right)^{-\frac{1}{\alpha}}$ 
and $\Gamma(\cdot)$ is the Gamma function. 
With Eq.~(\ref{eq:asss}) we can write 
\[
\P\left(\frac{\Ain(t)-rt}{b t^{H}}>\sigma\right)
\sim \left(c_{\alpha}\sigma\right)^{-\alpha}~, ~\sigma\rightarrow\infty~,
\]
Matching this expression with Eq.~(\ref{eq:envelope}) we
obtain the remaining parameter $K$ of the  \htss envelope by setting  
\begin{equation}
K = 
\left(\frac{b}{c_\alpha}\right)^{\alpha}~.\label{eq:envAlpha1}
\end{equation}
By Eq.~(\ref{eq:approxTailStable}), this 
envelope only holds for large $\sigma$, 
or, equivalently, low violation probabilities. 
An alternative method to obtain an \htss envelope for {\em all} values of $\sigma$ 
is to take advantage of the quantiles of $S_\alpha$. Since the density of $S_{\alpha}$ is not available 
in a closed form, the quantiles must be obtained 
numerically or by a table lookup. 
Let the 
quantile $z(\eps)$ be the value satisfying
\begin{equation}
P\big(S_{\alpha}>z(\eps)\big)=\eps~.
\label{eq:quantile}
\end{equation}
We obtain a statistical envelope by setting 
$\G(t; \sigma) = rt+z(\eps)\sigma t^{H}$ with a fixed $\eps$. 
In fact, this is the envelope for $\alpha$-stable processes 
from \cite{Barbosa05}. However, since $z(\eps)$ does not follow a 
power law, it is not an \htss envelope.
To obtain an \htss envelope from the quantiles, 
we express Eq.~(\ref{eq:envelope}) in terms of 
Eq.~(\ref{eq:quantile}), which can be done by setting 
\begin{equation}
K = \sup_{0<\eps<1}\left\{\eps \cdot \left(b 
z(\eps)\right)^{\alpha}\right\}~.\label{eq:envAlpha1q}
\end{equation}
In Fig.~\ref{fig:alpha}, we present  envelopes for a process satisfying 
Eq.~(\ref{eq:asss}) with 
\[
r= 75~Mbps ,~~\alpha= 1.6,~~H= 0.8,~~b =  60~Mbps ~ . 
\]
We show statistical envelopes with fixed violation probabilities 
$\eps = 10^{-1}, 10^{-2}, 10^{-3}$. 
The graph compares the \htss envelopes constructed with the tail approximation 
using Eq.~(\ref{eq:approxTailStable}) to those 
obtained from the quantiles via Eq.~(\ref{eq:envAlpha1q}). 
As can be expected, the envelopes computed from the asymptotic 
tail approximation are smaller than the quantile envelopes. 
We add that envelopes computed from the quantiles for a fixed $\eps$, 
as described in Eq.~(\ref{eq:quantile}), are very close to the 
tail approximation envelopes. 
If the corresponding envelopes were included in the figure, 
they would appear almost indistinguishable, 
suggesting that Eq.~(\ref{eq:approxTailStable}) 
provides reasonable bounds for all values of $\sigma$.

\subsection{Pareto Packet Distribution}\label{sec:constructPareto}

As second case study, we present an \htss envelope construction for 
a packet source with a Pareto arrival distribution. 
Packets arrive evenly spaced at rate $\lambda$ and 
packet sizes are i.i.d. described by a Pareto random variable $X_i$ 
for the $i$-th packet with tail distribution 
\begin{equation}
\P\Big(X_i>x\Big)=\left(\frac{x}{b}\right)^{-\alpha},~~~x\geq b~,\label{eq:pareto}
\end{equation}
where $\alpha\in(1,2)$.  $X$ has finite mean
$E\left[X\right]=\frac{b\alpha}{\alpha-1}$ and infinite
variance.  We will construct an \htss envelope for
the compound arrival process
\begin{equation}
\Ain(t)=\sum_{i=1}^{N(t)}X_i~,\label{eq:compound}
\end{equation}
where $N(t)=\lfloor\lambda t\rfloor$ denotes the number of packets which arrive by time~$t$. 
This arrival process is asymptotically self-similar with a Hurst parameter 
of $H = 1/\alpha$.

For the \htss envelope construction of the Pareto source, we take advantage of 
the {\em generalized central limit theorem} (GCLT) 
\cite{Nolan09}, which states that the $\alpha$-stable distribution $S_\alpha$ appears as 
the limit of normalized sums of i.i.d. random variables.  
For the independent Pareto random variables $X_i$, the GCLT 
yields 
\begin{equation}
\frac{\sum_{i=1}^{n}X_i-nE\left[X\right]}{c_{\alpha}n^{\frac{1}{\alpha}}}
\stackrel{n \rightarrow  \infty}{\longrightarrow}
S_{\alpha} \, \label{eq:gclt1}
\end{equation}
in distribution. Since the GCLT is an asymptotic 
limit, envelopes derived with the GCLT 
are approximate, with higher accuracy for larger values of $n$. 
Using that $N(t) \approx \lambda t$  for suitable large values, 
we can write the arrival function in Eq.~(\ref{eq:compound}) with 
Eq.~(\ref{eq:gclt1}) as 
\[ %
\Ain(t) \approx \lambda t E\left[X\right] 
+ c_\alpha (\lambda t )^{1/\alpha}  S_{\alpha}~, %
\] %
Since this expression takes the same form as Eq.~(\ref{eq:asss}), 
we can now  use the tail estimate of  Eq.~(\ref{eq:approxTailStable}) to obtain  
an \htss envelope with parameters 
\begin{equation}
r=\lambda E[X],~~\alpha,~~H=\frac{1}{\alpha},~~K\approx \lambda~.\label{eq:envPareto1}
\end{equation}
The same parameters are valid when 
$N(t)$ is a Poisson process, according to Theorem~3.1 in
\cite{Mikosch97}. 

Similar techniques can yield \htss envelopes for other 
heavy-tailed processes. For example, an aggregation of 
independent On-Off periods, where the duration of `On' and `Off' periods is 
governed by independent Pareto random variables yields an 
$\alpha$-stable process \cite{Mikosch02} 
in the limit of many flows $(N\rightarrow \infty)$ 
and large time scales  $(t\rightarrow \infty)$. 
This aggregate process is particularly interesting since dependent 
on the order in which the limits of $N$ and $t$ are taken, 
one obtains processes that are self-similar, but not heavy-tailed 
(fractional Brownian motion), processes that are heavy-tailed, but not 
self-similar ($\alpha$-stable L\'{e}vy motion), or a 
general $\alpha$-stable process. An approximation by an $\alpha$-stable process followed by 
an estimation of \htss parameters can also be  
reproduced for the M/G/$\infty$ arrival model \cite{Mikosch02}.

\medskip
\noindent{\bf Example. } 
We next compare envelope constructions for a Pareto source 
with evenly spaced packet arrivals with a size distribution 
given by Eq.~(\ref{eq:pareto}). The parameters are
$$%
\alpha= 1.6 ,~~b= 150~Byte,~~\lambda = 75~Mbps ~.
$$%
With these values, the average packet size is $400$~Byte. 
We evaluate the following types of envelopes:
\begin{enumerate} 

\vspace{5pt}
\item {\em \htss GCLT envelope.} This refers to the envelope constructed with the GCLT according to Eq.~(\ref{eq:envPareto1}). The value of $\sigma$ of the \htss envelope is set so that the right hand side of Eq.~(\ref{eq:envelope}) satisfies a violation probability of 
$\eps =10^{-3}$.

\vspace{5pt}
\item {\em Deterministic trace envelope.} This envelope is computed from a simulation of a packet trace with 1~million packets drawn from the given Pareto distribution. 
We compute the smallest envelope for the trace that satisfies Eq.~(\ref{eq:localenv1}) with $\eps(\sigma) = 0$ for all $\sigma>0$. 
The deterministic trace envelope, which is computed by $\G(t) = \sup_{\tau} \{A(t+\tau) - A(\tau)\}$ \cite{Book-Leboudec}, 
provides the smallest envelope of a trace that is never violated. 

\vspace{5pt}
\item {\em \htss trace envelope.} This is an \htss envelope  created from the same Pareto packet trace. We assume that the values of $\alpha$ and $H= 1/\alpha$ 
are given, but that the distribution is not known. 
The envelope is created directly from  Eq.~(\ref{eq:envelope}) by inspecting the relative frequency at which subintervals of the trace violate the \htss envelope. First, $K$ is selected as the smallest number that satisfies the right hand side of Eq.~(\ref{eq:envelope}) for all values of~$\sigma$. Then 
$\sigma$ is found by fixing the violation probability $\eps = 10^{-3}$. 

\vspace{5pt}
\item {\em Average rate.} For reference, we also include the average rate of the data in the figures, 
which is obtained from the same packet trace as in the trace envelopes. 

\end{enumerate} 

\begin{figure}[hT]
\centerline{\includegraphics[width=0.4\textwidth]{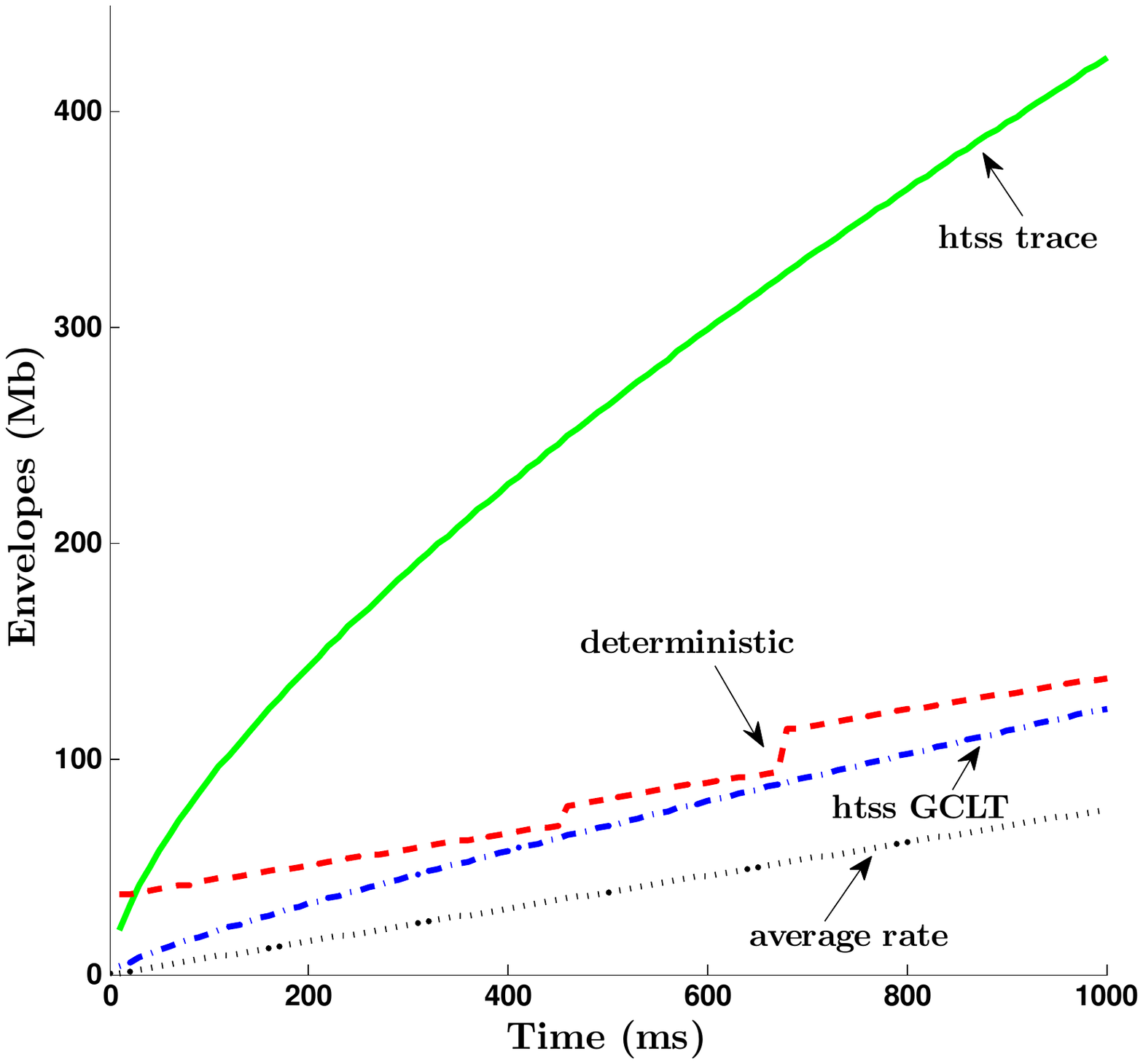}}
\caption{Envelopes for a Pareto packet source ($\eps=10^{-3}$).} \label{fig:DPareto}

\centerline{\includegraphics[width=0.4\textwidth]{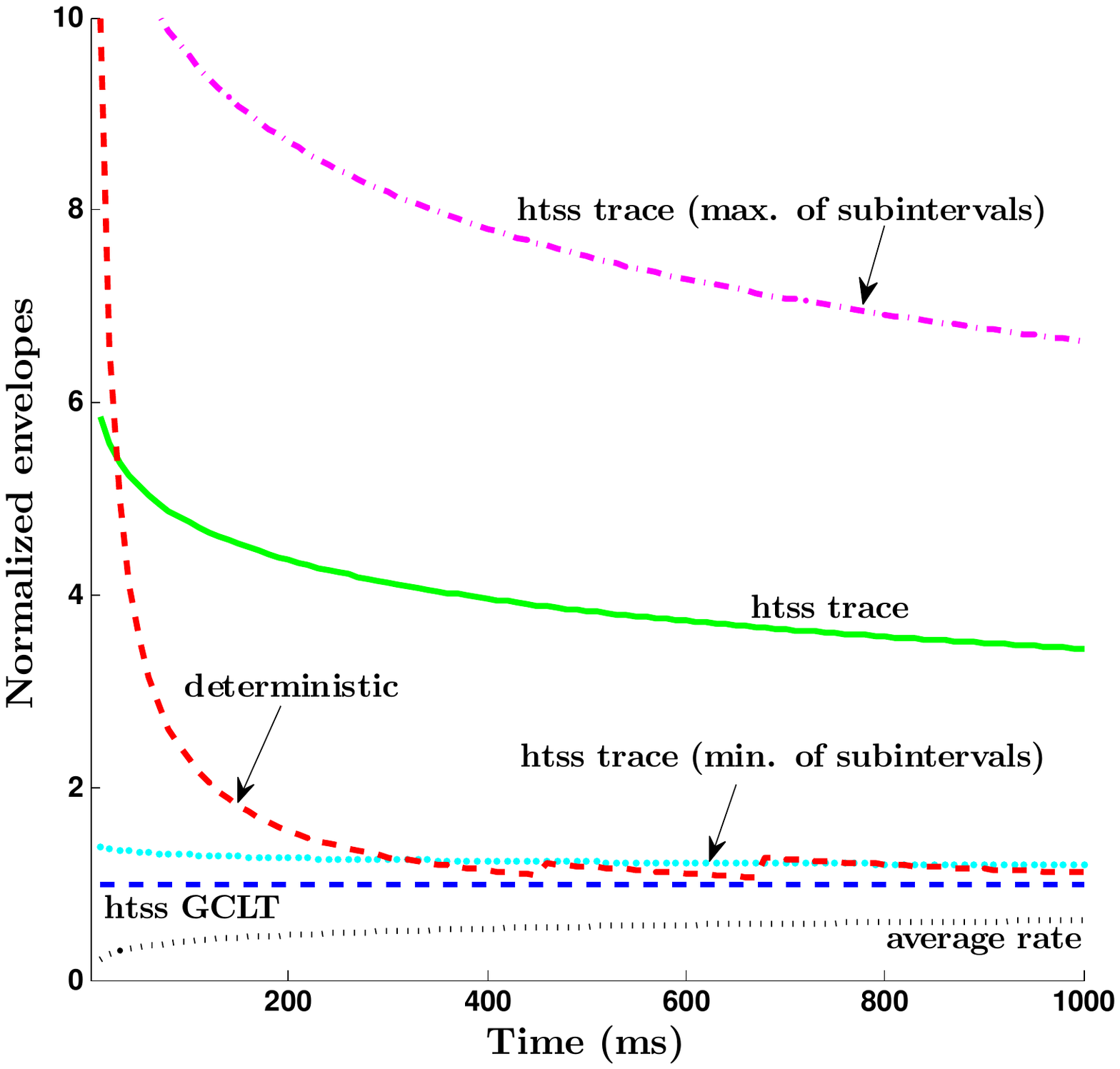}}
\caption{Normalized envelopes for a Pareto packet source ($\eps=10^{-3}$).} \label{fig:DPareto2}

\end{figure}

\medskip
The resulting envelopes are  plotted in Fig.~\ref{fig:DPareto}. The discrete steps of the deterministic trace envelope around $t = 0$~ms,  $t = 460$~ms, and $t=680$~ms are due to 
arrivals of very large  packets at certain times in the simulated trace. 
The \htss GCLT envelope is quite close to the data of the average rate. 
However, since the GCLT is an asymptotic result, this envelope is possibly too optimistic. 
On the other hand, the safely conservative \htss trace envelope 
is much larger than the corresponding deterministic trace envelope. 
The reason is that the construction of this envelope performs a heavy-tailed extrapolation 
of the data trace, and thus amplifies the variability of the underlying trace.

To investigate the variability of \htss trace envelopes as a function of the length of the 
trace, 
we present envelopes obtained from subintervals of the trace used for 
Fig.~\ref{fig:DPareto}. We use 
non-overlapping subintervals of 100,000 packets from the trace and compute \htss trace envelopes for the subintervals. In Fig.~\ref{fig:DPareto2}, we show the envelopes normalized by the values of the \htss GCLT envelope. We 
plot the maximum and the minimum values of the computed \htss envelopes for the subintervals. 
The figure makes drastically clear that the range of values of the \htss envelopes 
for the shorter intervals cover a wide range. This illustrates 
an inherent problem with generating a traffic characterization for heavy-tailed traffic  
from limited data sets. 
 
\begin{figure}[hT]
\centerline{\includegraphics[width=0.4\textwidth]{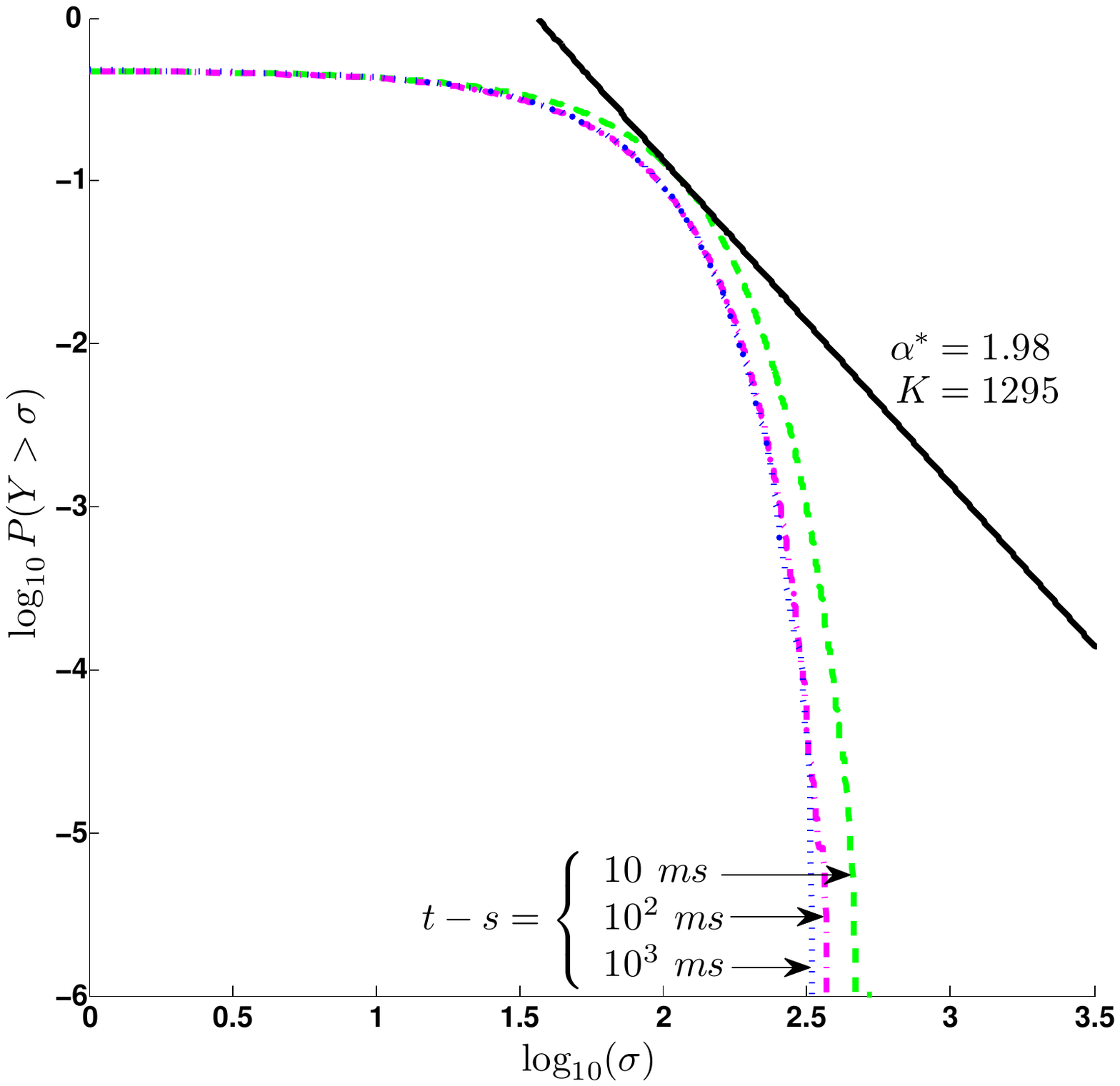}}
\caption{Normalized log-log plot of {\em Munich} trace.} \label{fig:traceloglog}

\centerline{\includegraphics[width=0.4\textwidth]{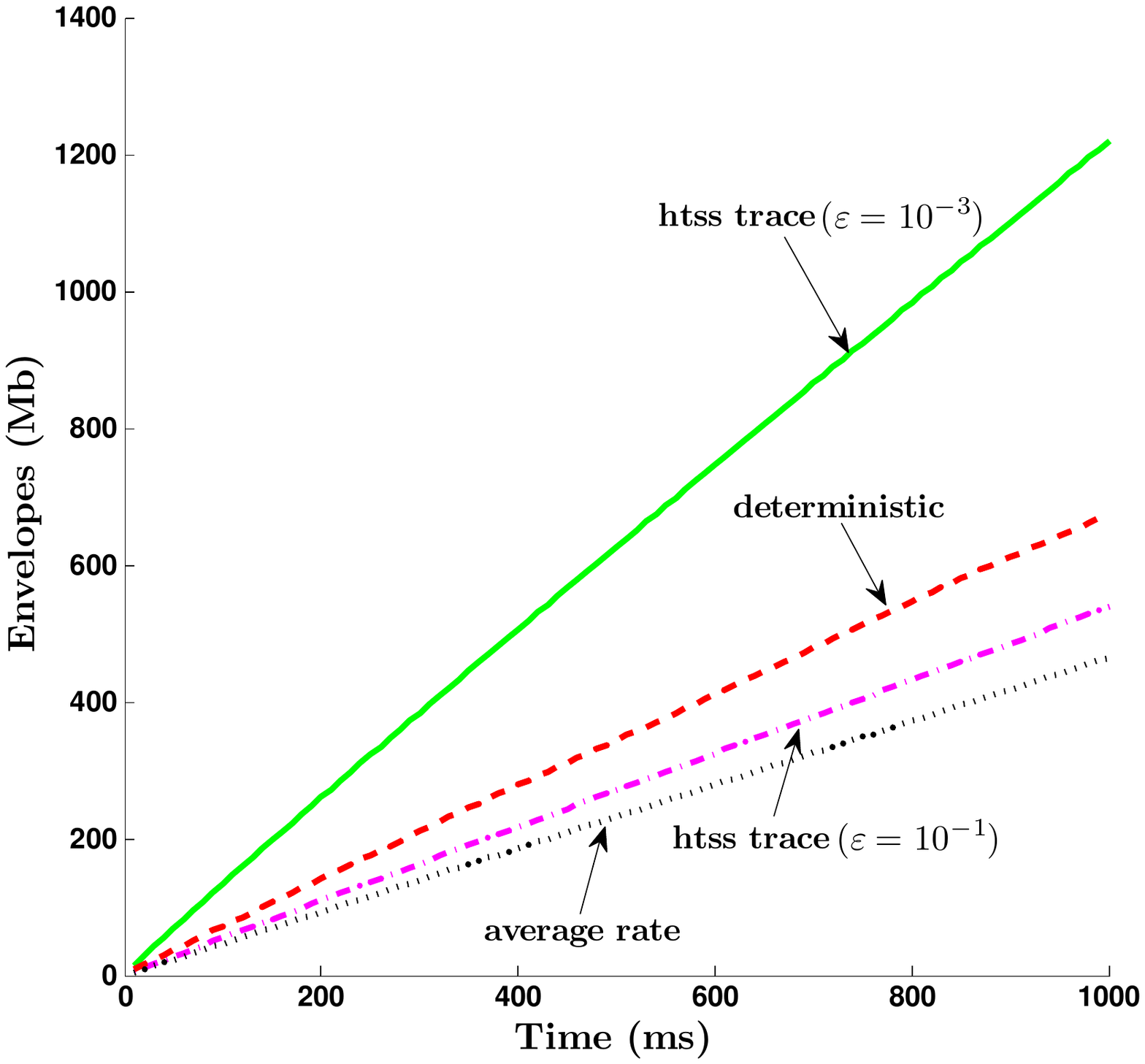}}
\caption{\htss envelopes for {\em Munich} trace.} \label{fig:traceenv}

\end{figure}
\subsection{Measured Packet Traces}\label{sec:traces}

We next show how to obtain an \htss envelope 
from measured traffic traces. 
The trace data was collected in October 2005 at the 1~Gbps uplink of the Munich Scientific 
Network, a network  with more than 50,000~hosts, 
to the German research  backbone network. 
The complete trace contains more than 6~billion packets, collected 
over a 24-hour time period.  Further details on the 
data trace and the collection methodology can be found in \cite{munich06}.
From this data, we 
select the first $10^9$~packets corresponding to  $2.75$~hours worth 
of data, with an overall average rate of $r^* = 465$~Mbps.  
We refer to this data as the {\em Munich} trace. 

To extract the tail index  and the Hurst parameter from the trace, we take 
advantage of parameter estimation methods for stable processes 
from \cite{McCulloch1986},\footnote{We use source code provided to us 
by the authors of \cite{Bates97theestimation}.}
which yields the following parameters for the {\em Munich} trace: 
\[
\alpha^* = 1.98 \ , ~~~~ 
H^*= 0.93  ~ .
\]
The remaining parameter $K$ needed for the $\htss$ envelope can now be obtained 
in the same way as described for the $\htss$ trace envelope.

To provide a sense of the trace data, we present in Fig.~\ref{fig:traceloglog} 
a normalized log-log plot of the {\em Munich} 
trace data in terms of the  normalized random variable
\[ %
Y:= \frac{\Ain(s,t)- r^*(t-s)}{(t-s)^{H^*}} ~.\label{eq:normalized}
\] %
Since  
$\P (Y > \sigma) = \P (\Ain(s,t) > \G (t-s;\sigma))$ 
where $\G$ is given in Eq.~(\ref{eq:htts}), the distribution of 
$Y$ corresponds to that of violations of the \htss envelope.  
In the figure, we show the log-log plot of $Y$ for different values of $(t-s)$, namely, 
$t-s =  10, 100, 1000$~ms.  
If the trace data was self-similar with the exact Hurst parameter $H^*$, the log-log 
data curves should 
match perfectly for all values of  $(t-s)$. (We note that by reducing the value of the Hurst parameter 
slightly, the curves for different values of $(t-s)$ can be made to match up almost 
perfectly). 
Since the decay of the log-log plots is obviously not linear, 
the distribution of the {\em Munich} trace does not appear to be heavy-tailed. 
We will see that a characterization of such a  non-heavy-tailed process by an 
\htss envelope leads to a pessimistic estimation. 

We can also use Fig.~\ref{fig:traceloglog} to graphically construct an \htss envelope 
for the {\em Munich} trace. 
Since we already have determined the tail index $\alpha^*$ and the Hurst parameter $H^*$ as given above, 
we only need to find~$K$. 
The value of this parameter can be obtained by taking the logarithm of Eq.~(\ref{eq:envelope}). Using 
the definition of $Y$, this yields
\[
\log \P\Big( Y > \sigma \Big)\leq \log K - \alpha^* \log \sigma ~.
\]
Applying this relationship to Fig.~\ref{fig:traceloglog}, we should select $K$ as the smallest value 
such that the linear function $\log K - \alpha^* \log \sigma$ lies above  the log-log plots of  $\P( Y > \sigma )$ 
in the figure. 
In Fig.~\ref{fig:traceloglog}, we include the linear segment with $K=1225$ as a thick line. 
Clearly, any other selection of $K$ and $\alpha^*$ providing an upper bound of the 
log-log plots of the Munich trace also yields a valid \htss envelope for all values of $\sigma$. 
An \htss  envelope for a fixed 
violation probability $\eps$ can be obtained from Fig.~\ref{fig:traceloglog} by 
finding the value of $\sigma$ that corresponds to the desired violation probability 
of  the linear segment. 
Finally, we can use Fig.~\ref{fig:traceloglog} to assess the accuracy of the \htss envelope. 
The linear segment (the thick black line) is close to the trace data when $ \P( Y > \sigma)\approx 10^{-1}$. 
Otherwise, the linear segment is quite far apart 
from the plots of the trace. This indicates that the \htss envelopes developed with the parameter settings 
used for the linear segment are accurate only when the violation probability is around $10^{-1}$.
If the data trace was truly heavy-tailed, 
the data curves would maintain a linear rate of decline at a rate around $\alpha^*$, and would 
remain close to the linear segment for any $\sigma$ sufficiently large. 
  
In Fig.~\ref{fig:traceenv}, we show \htss envelopes for the {\em Munich} trace obtained 
with  the linear segment from Fig.~\ref{fig:traceloglog} for  $\eps = 10^{-1}$ and $\eps = 10^{-3}$. 
For comparison, we include in Fig.~\ref{fig:traceenv} the average rate of 
the traffic trace, as well as a deterministic envelopes of the 1~sec subinterval of the 
trace that generates the most traffic. 
(Since the computation of a deterministic  trace envelope as defined in 
Subsection~\ref{sec:constructPareto} grows quadratically in the size of the trace,
the computation time to construct a deterministic envelope for the complete {\em Munich}  trace is 
prohibitive. The included deterministic envelope for a subinterval of the trace
is a lower bound for the deterministic envelope of the complete trace. However, 
for the depicted time intervals, the deterministic envelope for the subinterval is a good  
representation of the deterministic envelope of the entire trace, for several reasons. 
First, by selection of the subinterval, 
at $t=1000$~ms the envelope of the subinterval  and the envelope of the complete 
trace are identical. 
Second, since any deterministic envelope is a subadditive function, the slope 
of the envelope decreases for larger values of time. 
Now, any function that satisfies these properties 
cannot vary significantly from the depicted envelope of the selected subinterval.)
Comparing the \htss envelopes with the reference curves confirms our earlier discussion on the accuracy 
of the \htss envelopes: For  $\eps=10^{-1}$, the \htss envelope is close to  
the plot of the average rate. On the other hand, the envelope for $\eps = 10^{-3}$ is 
quite pessimistic, and lies well above the deterministic envelope. 

\section{Service Guarantees with Heavy Tails} 
\label{sec:service}

We next formulate service guarantees with a power-law decay.  
In the network calculus, 
service guarantees are expressed in terms of functions 
that express for  a given arrival function a 
lower bound on the departures. 
In general, a {\em statistical service curve} is
a function $\S(t;\sigma)$ such that for all 
$t \geq 0$ and for all $\sigma > 0$ 
$$
\P\bigl( \Aout(t) < \Ain\conv \S(t;\sigma) \bigr)
\le \eps(\sigma)\, . 
$$
Here, 
$$
\Ain\conv \S(t;\sigma)=
\inf_{s\le t}\bigl\{\Ain(s)+\S(t-s;\sigma)\bigr\}
$$
denotes the min-plus convolution of the arrivals with the
service curve $\S(t;\sigma)$,
and $\eps$ is a non-increasing function 
that satisfies $\eps(\sigma)\to 0$ as $\sigma\to\infty$.

We define a {\em heavy-tailed (ht) service curve} as a service curve
of the form
\begin{equation}
\S(t,\sigma)=[Rt-\sigma]_+\,,\quad
\eps(\sigma) =L\sigma^{-\beta}
\label{eq:htt}
\end{equation}
for some $\beta$ with $0<\beta<2$ and some constant $L$.
In analogy to the formulation of traffic envelopes 
in Section~\ref{sec:envelope}, the \htt 
service curve specifies that the deviation from the 
service rate guarantee $R$ has a heavy-tailed decay. 
The rationale for not including a Hurst parameter 
in the definition of the \htt service guarantees is that 
the form of Eq.~(\ref{eq:htt}) facilitates the computation of 
service bounds over multiple nodes. 
In this paper, we consider two types of \htt service curves, one characterizing the 
available capacity at a link with cross-traffic, the other modeling a packetizer. 

\bigskip
\noindent
$\bullet$~{\it Service at link with cross traffic (Leftover Service):} 
This service curve seeks to describe the  
service available to a selected flow at a constant-rate link with capacity $C$, 
where the competing traffic at the link, referred as {\em cross traffic}, 
is given by an \htss envelope.
By considering the pessimistic case that the selected flow receives 
a lower priority than the cross traffic, we will obtain  
a lower bound for the service guarantees for most 
workconserving multiplexers \cite{Book-Leboudec}. 
Since the service guarantee  of the selected flow 
consists of the capacity that is left unused by 
cross traffic,  we refer to the service interpretation  
as  {\em leftover service}. 
Since the derivation of an \htt service curve for such a leftover service 
requires a sample path bound for the \htss cross traffic, we defer the derivation 
to Subsection~\ref{subsec-leftover}.

\bigskip
\noindent
$\bullet$~{\it Packetizer:} 
We will also use the \htt service model  
to express a packetized view of traffic  
with a heavy-tailed packet size distribution. We model 
discrete packet sizes by 
a service element that delays traffic until all bits  belonging to 
the same packet have arrived, and then releases all bits of the packet at once. 
Such an element is referred to as a {\em packetizer}.  
By investigating packetized traffic we can relate our bounds 
to a queuing theoretic analysis with 
a packet-level interpretation of traffic (see Section~\ref{sec:scaling}). 
We now derive a service curve for a packetizer. 
For a packet-size distribution satisfying 
$\P\bigl( X>\sigma\bigr\}\le L\sigma^{-\alpha}$,
we show that a
constant-rate workconserving link
of capacity $C$ provides an \htt service curve with
rate $R=C$ and a suitable function $\eps(\sigma)$. 

Denote by $X^*(t)$ the part of the 
packet in transmission at time $t$ that has 
already been transmitted.
We can view $X^{*}(t)$ as the
current lifetime of a renewal process. 
It is known from the theory of renewal processes 
(see \cite{Karlin75}, pp. 194) that 
\begin{align*}
\lim_{t\rightarrow\infty}\P\Big(X^*(t)> \sigma\Big)
& =\frac{\int_{\sigma}^{\infty} \P(X>x) \, dx}{E[X]} \,.
\end{align*}
The bound on the tail probability holds for all times $t$, provided 
that the arrival time of the first packet
after the network is started with empty queues at
$t=0$ is properly randomized.
Using $X^*(t)$, the departures of a packetizer are given by 
\[
D(t) = \left\{
\begin{array}{l l}
A(t) \ , & \underline{t} = t \ , \\
A(\underline{t}) + C (t - \underline{t}) - X^* (t) \ , & \underline{t} < t \ , 
\end{array}
\right. 
\]
where $\underline{t}$ is the beginning of the busy period of $t$. 
Set $\S(t;\sigma)=[Ct-\sigma]_+$. 
If $\rho$ is the 
utilization of $A$ as a fraction of the link rate
$C$, that is, 
\[
\rho = \sup_{s\leq t}\frac{E[A(s,t)]}{C(t-s)} \, ,
\] 
then 
\begin{align}
\notag
\P\Big(\Aout(t)< \Ain \conv \S(t;\sigma) \Big)
& \leq  \rho\P\Big(X^{*}(t)>\sigma\Big) \notag\\
&\le \frac{\rho L}{(\alpha-1) E[X]} \sigma^{-(\alpha-1)}\,.
\label{eq:packetservice}
\end{align}

\begin{figure}
\centerline{\includegraphics[width=0.6\textwidth]{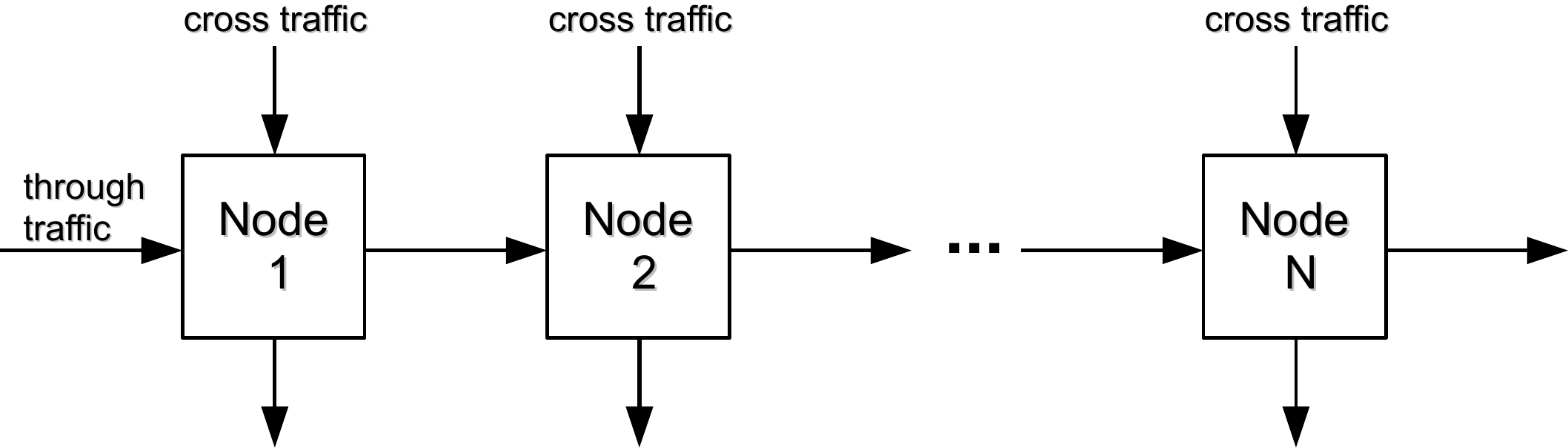}}
\caption{A network with cross traffic.} \label{fig:crossTraffic}
\end{figure}

\section{Network Calculus with {\em htss} Envelopes}
\label{sec:convolution}

We consider a network as in Fig.~\ref{fig:crossTraffic}.
A flow traverses $N$ nodes in series. Its traffic is referred to as 
through traffic.
At each node, the through traffic is multiplexed with 
arrivals from competing flows, called \textit{cross traffic}. 
Both through and cross traffic are described by \htss envelopes.
We will present results that  yield bounds on 
the end-to-end delay and  backlog of the through traffic.

\subsection{Statistical sample path Envelope} 

The network calculus for heavy-tailed traffic is enabled by a statistical 
sample path envelope for traffic with \htss envelopes. 
To  motivate the relevance of the sample path bound,
let us consider the backlog of a flow at 
a workconserving link that operates
at a constant rate $C$. The backlog at time $t$ 
is given by
$$
B(t)=\sup_{s\le t} \bigl\{ \Ain(s,t)-C(t-s)\bigr\}\,.  $$
Notice that the backlog expression depends on the entire
arrival sample path $\left\{\Ain(s,t)\right\}_{s\leq t}$.
To compute an upper bound
for the tail probability $\P\bigl(B(t)>\sigma\bigr)$, 
in many places in the literature, 
in particular, in all prior works attempting a 
network calculus analysis with 
heavy-tailed traffic \cite{Makra2000,Barbosa05,Jiang05a,Hatzinakos01}, 
the tail distribution is approximated by
$$
\P  \bigl(B(t)>\sigma\bigr)
\approx \sup_{s\le t} 
\P \bigl( \Ain(s,t)-C(t-s)>\sigma \bigr) \,.
$$
However, the right hand side is
generally {\em smaller} than the left hand side. 
Applying to the right hand side 
a statistical envelope that only
satisfies Eq.~(\ref{eq:localenv1}) but not Eq.~(\ref{eq:env1})  
does not yield an upper bound
but rather an upper bound to a lower bound.
The derivation of rigorous upper bounds requires 
a sample path bound for the arrivals. To derive 
such bounds, we discretize time by setting
$x_k=\tau \gamma^k$, where $\tau>0$ and $\gamma>1$ are
constants that will be chosen below. If 
$t-x_k\le s< t-x_{k-1}$, then
$$
\Ain(s,t)-C(t-s) \le \Ain(t-x_k,t)-C x_{k-1} \,.
$$
It follows that 
$$B(t)\le \sup_{k} \bigl\{ \Ain(t-x_k,t)-C x_{k-1}\bigr\}\,. $$
If the arrivals satisfy an \htss envelope
$\G(t)=rt+\sigma t^H$ with 
$\eps(\sigma)=K\sigma^{-\alpha}$, we obtain
with the union bound 
\begin{eqnarray}
\P\bigl(B(t)>\sigma\bigr)
&\le&\sum_{k=-\infty}^\infty 
\P\bigl(\Ain(t-x_k, t)> \sigma + C x_{k-1}\bigr)  \nonumber\\
& \le & \frac{1}{H(1-H) \log \gamma} \int_{z}^{\infty} K x^{-\alpha-1} dx \Big\vert_{z=\frac{(C/\gamma -r)^H \sigma^{1-H}}{\gamma^{H(1-H)}}} \nonumber\\
&\le& \tilde K \sigma^{-\alpha(1-H)}\,. \label{eq:def-constant-0}
\label{eq:B-rigorous}
\end{eqnarray}
In the second  line we have used
Lemma~\ref{lm:est1} from the appendix 
to evaluate the sum.
Writing  $C=r+ \mu$ and minimizing over $\gamma$ gives the constant 
\begin{equation}\label{eq:def-constant}
\tilde K= K \cdot  \inf_{1<\gamma<1+\frac{\mu}{r}} 
\Bigl\{\Bigl(\frac{r+\mu}{\gamma}-r\Bigr)^{-\alpha H} 
\frac{\gamma^{\alpha H(1-H)} }
{\alpha H(1-H)\log\gamma}\Bigr\} \, .
\end{equation}
We remark that, typically, we have $1<\alpha<2$ 
and $\alpha^{-1}\le H<1$, so that $\alpha(1-H)<1$. This means that 
the backlog is almost surely finite, but cannot be expected
to have finite mean.

The main technical ingredient of the above proof of the backlog 
bound is the discretization of time by the 
geometric sequence $x_k=\gamma^k\tau$.
This is an instance of
\textit{under-sampling}, where not every time step is
used in probabilistic estimates. Ä
Commonly in the literature, 
time is discretized by dividing it into equal units
with $x_k=k\tau$. 
In \cite{vojnovic03tc}, the choice is described as a
general optimization problem over
arbitrary sequences $x_0\leq x_1\leq \ldots\leq t$, but
not applied, since all examples in \cite{vojnovic03tc} 
only optimize over $\tau$ in uniformly spaced sequences.
Note that using a uniform discretization in 
the derivation of Eq.~(\ref{eq:B-rigorous}) 
would cause the infinite sum to become unbounded.  

An immediate consequence of the backlog 
bound is a sample path bound for \htss envelopes.

\begin{lemma} {\sc \htt sample path Envelope. } 
\label{lm:arrival-samplepath}
If arrivals to a flow are bounded by an \htss envelope
$$
\G(t;\sigma)=rt+\sigma t^H\,,\quad \eps(\sigma)=K\sigma^{-\alpha} \ , 
$$
then, for every choice of $\mu>0$,  
$$
\overline{\G}(t;\sigma)=(r+\mu)t+\sigma\,,\quad
\overline{\eps}(\sigma)= \tilde K \sigma^{-\alpha(1-H)}\,,
$$
is a statistical sample path envelope according to
Eq.~(\ref{eq:env1}).  The constant $\tilde K$ 
is given by Eq.~(\ref{eq:def-constant}). 
\end{lemma}

The proof follows immediately from 
Eq.~(\ref{eq:B-rigorous}) by replacing $C$ with 
the relaxed arrival rate $r+\mu$. 
The \htt sample path envelope is reminiscent of a
leaky-bucket constraint with a single burst and rate, and does
not reflect the self-similar scaling of the
\htss envelope.

We note that a small modification of the proof 
would yield a sample path envelope of the form
$$
\overline{\G}(t,\sigma)=(r+\mu)t+\sigma t^H+M\,,\quad
\overline{\eps}(\sigma)= L \sigma^{-\alpha} \,,
$$
which retains the self-similar scaling properties
of the \htss envelope. The constant $L$
depends on the parameters $\alpha, H, r, \mu$ and
on the choice of $M>0$.
The reason we prefer the simpler envelope given
by Lemma~\ref{lm:arrival-samplepath}
is that it facilitates the estimation of the service provided
to a flow across multiple nodes.

\noindent 
\subsection{Heavy-Tailed Leftover Service Curve}
\label{subsec-leftover}
\begin{figure}[h]
\centerline{\includegraphics[width=0.4\textwidth]{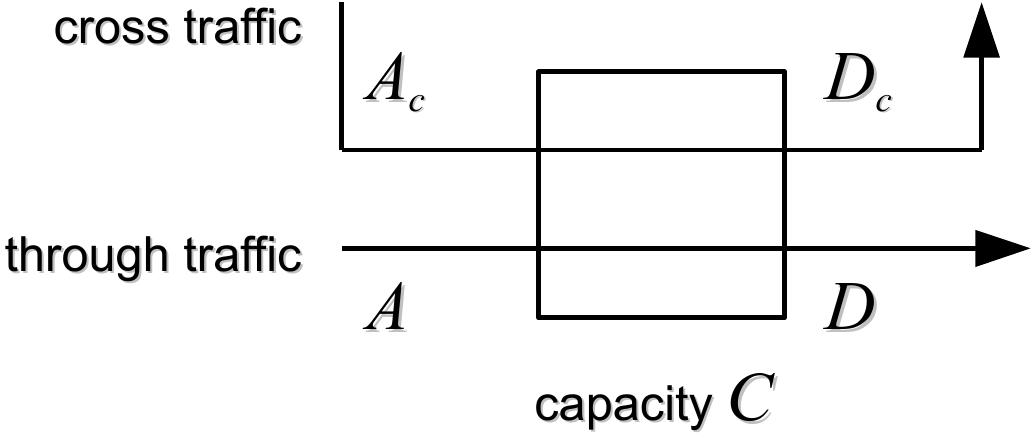}}
\caption{Constant-rate link with capacity $C$.} \label{fig:crossTraffic-single}
\end{figure}

With a sample path  envelope for heavy-tailed traffic at hand, 
we can now derive  a service curve for the heavy-tailed leftover service from 
Section~\ref{sec:service} at a constant-rate link with capacity $C$, as illustrated in 
Fig.~\ref{fig:crossTraffic-single}. 
We denote arrivals of the through flow by $\Ain$
and cross traffic arrivals by $\Ain_c$. 
Departures are denoted by $\Aout$ and $\Aout_c$, respectively.
Assuming that $\Ain_c$ is characterized by an 
\htss envelope of the form 
$\G_c (t) =  r_c(t-s)+\sigma(t-s)^{H_c}$ 
with $\eps(\sigma)=K_c\sigma^{-\alpha_c}$
where the bound on the arrival rate satisfies $r_c<C$, we will show that 
the through flow is guaranteed a \htt service curve
$\S(t;\sigma)=[Rt-\sigma]_+$ 
with rate $R=C-r_c-\mu$, and a violation probability
$\eps(\sigma)$ that can be estimated explicitly.
Here, $\mu$ is a free parameter.

Let $\underline{t}$
be the beginning of the busy period of $t$ at the link. 
Then, the aggregate departures in $[\underline{t},t)$ satisfy
$(\Aout + \Aout_c) (\underline{t},t)
= C(t-\underline{t})$, and departures for the cross
traffic satisfy
$\Aout_c(\underline{t}, t) \le \min\{ C(t-\underline{t}),
\Ain_c(t)-\Ain_c(\underline{t})\}$. With this we can derive 
\begin{eqnarray*}
\Aout(t)
&\ge&\Ain(\underline{t}) + 
\bigl[ C(t-\underline{t})-\Ain_c(\underline{t},t)\bigr]_+\\
&\ge& \inf_{s\le t} \bigl\{\Ain(s) + (C-r_c-\mu)(t-s) \bigr\} 
 - \sup_{s\le t}\{ \Ain_c(s, t)-(r_c+\mu)(t-s)\} \, , 
\end{eqnarray*}
for every choice of $\mu > 0$. We obtain 
\begin{eqnarray}
\P\bigl(\Aout(t)<\Ain \conv \S (t;\sigma) \bigr) 
&\le& \P\bigl(\sup_{s\le t}\{
\Ain_c(s,t)-(r_c+\mu)(t-s)\}>\sigma\bigr)\ \, \nonumber \\ %
&\le& \tilde K_c \sigma^{-\alpha_c(1-H_c)}\, ,
\label{eq:leftoverservice}
\end{eqnarray}
where $\tilde K_c$
is given by Eq.~(\ref{eq:def-constant})
This proves that $\S(t;\sigma)=[Rt-\sigma]_+$ is an \htt service curve. 

\smallskip
The description of the leftover service in Eq.~(\ref{eq:leftoverservice})
can be combined with Eq.~(\ref{eq:packetservice})
to characterize the leftover service available
to a packetized through flow at a node.
The result (which we state without proof)
is that at a link that
operates at rate $C>r_c$, the through flow
receives a service guarantee given by the \htt service curve
\begin{equation}
\label{eq:leftoverservicegeneral}
\S(t;\sigma)=[(C-r_c-\mu)t-\sigma]_+\,,\quad \eps(\sigma)=L\sigma^{-\beta}\,,
\end{equation}
where $\beta=\min\{\alpha_p-1,\alpha_c(1-H_c)\}$, 
$\alpha_p$ is the tail decay in Eq.~(\ref{eq:packetservice}), 
and $\mu>0$ is a free parameter. The violation probability is given by
$$
\eps(\sigma)=\inf_{\sigma_1+\sigma_2=\sigma} 
\left\{  \tilde K_c \sigma_1^{-\alpha_c(1-H) }
+\frac{\rho L_p}{(\alpha_p-1)\E[X]}\sigma_2^{-(\alpha_p-1)}
\right\}\,,
$$
where $\rho\le 1$ is the utilization of the through traffic as a fraction of $C$,
$E[X]$ is the average packet size,
and the constant 
$\tilde K_c$ is defined by Eq.~(\ref{eq:def-constant}) with $\tilde K_c$ 
in place of $K$. When traffic is not packetized, the second term in the sum above 
is equal to zero. 
The constant in Eq.~(\ref{eq:leftoverservicegeneral})
can be computed explicitly by first 
using Lemma~\ref{lemma:K-sigma-alpha} (Eq.~(\ref{eq:lower-power}))
to lower the larger exponent to $\beta$, and then
applying Lemma~\ref{lemma:K-sigma-alpha} (Eq.~(\ref{eq:minimize-sum})).

\subsection{Single Node Delay Analysis} 
We next present a delay bound  at a single node where arrivals 
are described by \htss envelopes and service is described by 
an \htt service curve. 

\begin{theorem}{\sc Single Node Delay Bound. }
\label{th:ppb} 
Consider a flow that is characterized by an \htss envelope 
with 
$\G(t,\sigma)=rt+\sigma(t-s)^{H}$ and $\eps(\sigma)=K \sigma^{-\alpha}$,
and  that receives an \htt service curve at a node given
by 
$\S(t;\sigma)=[Rt-\sigma]_+$ and $\eps(\sigma)=L \sigma^{-\beta}$.
If $r<R$, then the delay $W$ satisfies
\begin{eqnarray*}
\P\bigl(W(t)>w\bigr) &\le& M(Rw)^{-\min\{\alpha(1-H),\beta\}} \, ,
\end{eqnarray*}
where $M$ is a constant that depends on
$\alpha$, $H$,  $r$, $\mu=R-r$, and
$\beta$.
\end{theorem}

\bigskip\noindent 
{\sc Proof.}  Let $\Ain(t)$ and 
$\Aout(t)$ denote the arrival and departures of the (tagged) 
flow at the node.  The delay is given by
$$
W(t)=\inf\bigl\{t-s\ \mid\ \Ain(s)\le\Aout(t)\bigr\}\,.
$$
Fix $\sigma_1, \sigma_2>0$ with $R(\sigma_1+\sigma_2)=w$.
Suppose that on a particular sample path, 
$$
\sup_{s\le t-w}\bigl\{
\Ain(s,t-w) - R(t-s-w)\}\le \sigma_1\,,
$$
and that 
$$
D(t) \ge \inf_{s\le t} \bigl\{
  \Ain(s) + [R(t-s)-\sigma_2]_+\bigr\}\,.
$$
If the infimum is assumed for some $s\le t-w$, then
\begin{eqnarray*}
\Aout(t)&\ge& 
\Ain(s) + R(t-s)-\sigma_2\\
&\geq& \Ain(t-w)\, .
\end{eqnarray*}
If, on the other hand,
the infimum is assumed for some $s\ge t-w$, then
$$D(t)\ge A(s) \ge A(t-w)$$
by monotonicity. In both cases,  we see that
$W(t)\le w$. 
It follows with union bound that
\begin{align}
\P\bigl(W(t)>w\bigr) 
& \le \P\bigl( \sup_{s\le t-w}\{
\Ain(s,t-w)  - R(t-s-w)\}> \sigma_1\bigr) \nonumber \\
& 
\hskip3cm
+ \P\bigl(\Aout(t)< \inf_{s\le t}[\Ain(s) + R(t-s)-\sigma_2]_+\bigr) 
\nonumber\\
& \le \tilde K \sigma_1^{-\alpha(1-H)} +L\sigma_2^{-\beta}\,,
\label{eq:ppb}
\end{align}
where $\tilde K$ is defined by Eq.~(\ref{eq:def-constant}). 
For the first term, we have used
the sample path bound in Lemma~\ref{lm:arrival-samplepath}
with $\mu=R-r$,
and for the second term we have used the
definition of \htt service curves. The proof is completed by
first lowering the larger of the two
exponents to $\beta'=\min\{\alpha(1-H),\beta\}$ 
using Lemma~\ref{lemma:K-sigma-alpha} (Eq.~(\ref{eq:lower-power})), 
and then minimizing explicitly
over the choice of $\sigma_1$ and $\sigma_2$ using Lemma~\ref{lemma:K-sigma-alpha} 
(Eq.~(\ref{eq:minimize-sum})).
For the constant, this yields the estimate
\begin{equation}
M\le \Big\{ \tilde K ^{\frac{\beta'}{(1+ \beta')
\alpha(1-H)}}
+ L_s^{\frac{\beta'}{(1+ \beta')\beta}}
\Bigr\}^{1+ \beta'}\,.
\label{eq:yy}
\end{equation}
\hfill$\Box$

\begin{figure}[t]

\centerline{\includegraphics[width=0.4\textwidth]{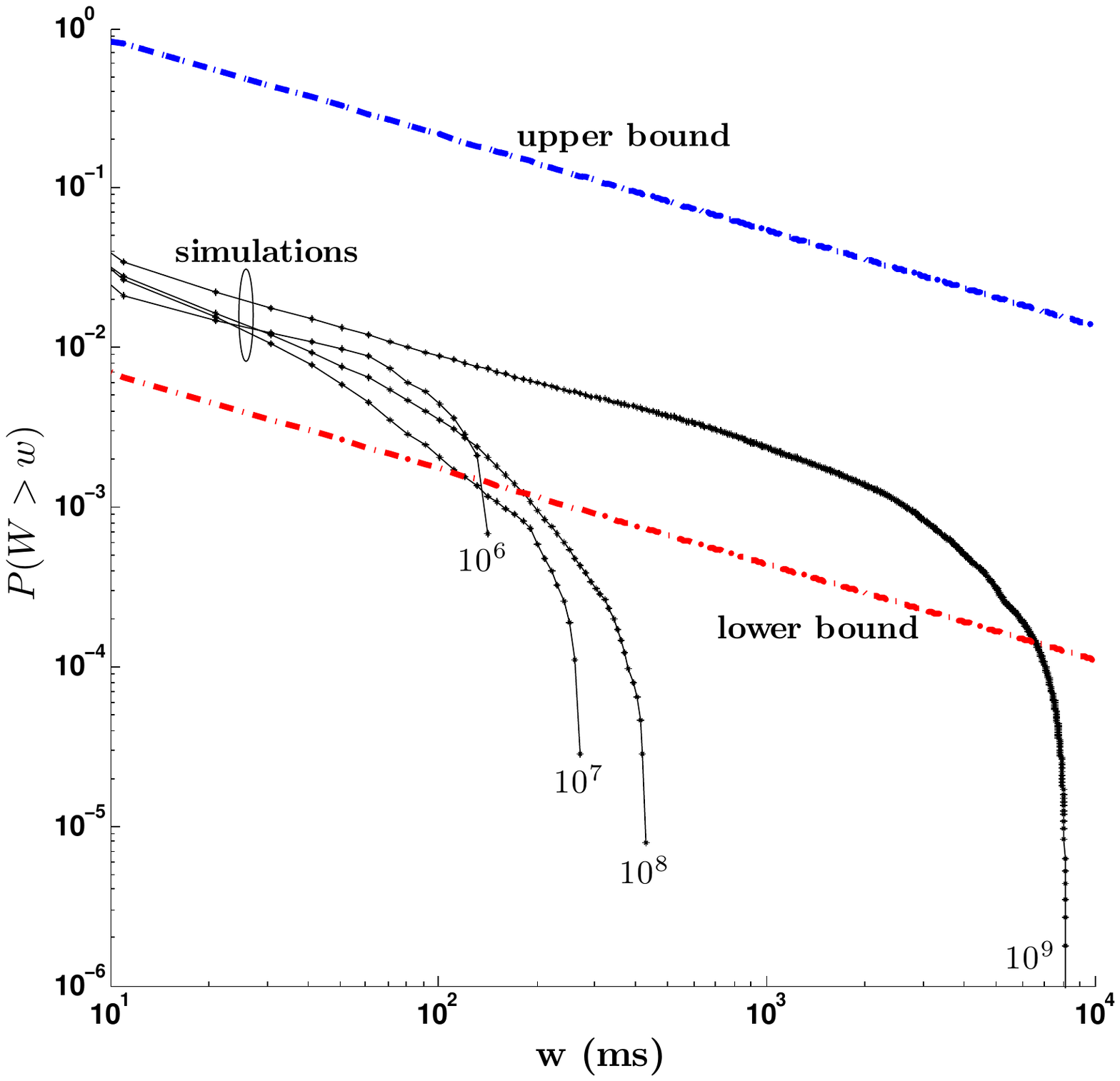}}

\caption{Log-log plot of single-node delays for a Pareto traffic source. Upper bounds and lower bounds are compared to simulation traces with $10^6,10^7,10^8$ and $10^9$ arrivals.} \label{fig:loglogdelay}
\end{figure}

\noindent{\bf Example: }
We compute the delay experienced by a Pareto traffic source 
at a 100~Mbps link.  The parameters are
$$%
\alpha= 1.6 ,~~b= 150~Byte,~~\lambda = 75~Mbps\, .
$$%
With these values, the average data unit has a size of $400$~Byte, 
and the link utilization is 75\%.
The service curve is computed from Eq.~(\ref{eq:packetservice}). The reason for 
selecting this example (which does not have cross traffic) is that 
it permits a comparison with a queueing theoretic result in \cite{BuLiCi06}, 
which presents a lower bound on the quantiles of a Pareto source in a tandem network with 
$N$ nodes and no cross traffic, $w_N(z)$,  as 
\begin{equation}
\label{eq:lower-quant-Pareto}
w_N(z) \ge \frac{ (N b )^{\frac{\alpha}{\alpha-1}}}
             {\bigl( (\alpha-1) \lambda^{-1}|\log( 1 - \eps)|\bigr)^{
\frac{1}{\alpha-1}}}\,.
\end{equation}
In Fig.~\ref{fig:loglogdelay} we show a log-log plot of the delay distribution. 
The graph illustrates the power-law decay for the upper bound and the lower bound from \cite{BuLiCi06}.
We also show the results of four simulation runs of an initially empty system 
with  $10^6, 10^7, 10^8$ and  $10^9$ packets. 
The simulation traces indicate that the actual delays may be closer to the lower bounds.
Note that the fidelity of the simulations deteriorates at smaller violation probabilities. 
Since even long simulations runs do not contain sufficiently many events with large delays, 
they violate analytical lower bounds. 
Even the simulation run of 1~billion arrivals does not maintain the 
power-law  decay for violation probabilities below $\eps =10^{-3}$.

\subsection{Multi-Node Delay Analysis}
\begin{figure}[ht]
\centerline{\includegraphics[width=0.5\textwidth]{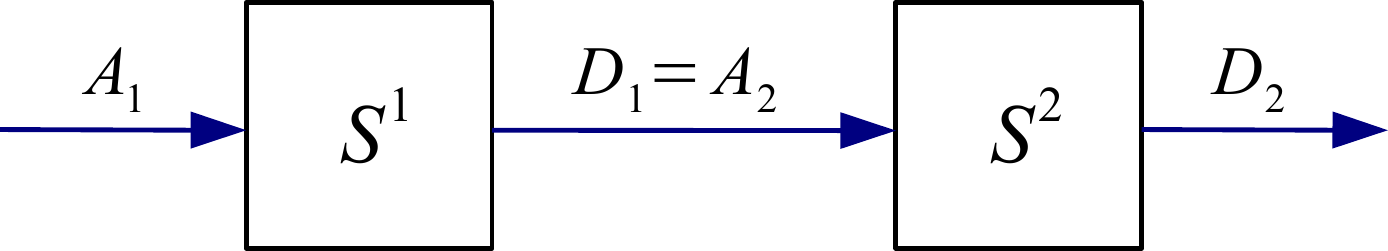}}
\caption{Two nodes in series.} \label{fig:network}
\end{figure}

We turn to the computation of end-to-end delays  for a complete 
network path.  As in the deterministic version of 
the network calculus  \cite{Book-Leboudec}
we express the service given by all nodes on the path  
in terms of a single service curve, and then 
apply single-node delay bounds. 
We start with a network of two nodes.
We denote by $A_1$ the arrivals 
of the analyzed flow at the first node, and by $D_1$ or $A_2$ 
the departures of the first node that arrive to the second node.

\bigskip 
\begin{lemma}{\sc Concatenation of two \htt service curves. }
\label{lm:convolution} Consider an arrival flow traversing
two nodes in series. 
The first node offers an \htt service curve with 
$\S_1(t,\sigma)=[R_1 t-\sigma]_+$ and $\eps_1(\sigma)=L_1 \sigma^{-\beta_1}$, 
and the second node offers a service 
curve $\S_2(t;\sigma)= [R_2t-\sigma]_+$ 
and an arbitrary function 
$\eps_2(\sigma)$.  Then for any $\gamma>1$,
the two nodes offer the
combined service curve given by 
\begin{align*}
& \S (t,\sigma) =  \Bigl[\min\Bigl\{R_1,\frac{R_2}{\gamma}\Bigr\} 
t-\sigma\Bigr]_+\ , \\
& \eps(\sigma) =  \inf_{\sigma_1+\sigma_2=\sigma}
\Bigl\{\tilde \eps_1(\sigma_1)
\bigl(|\log \tilde \eps_1(\sigma_1)|+2\bigr)2^{[\beta_1-1]_+}
+ \eps_2(\sigma_2)\Bigr\}\,,
\end{align*}
where $\tilde \eps_1(\sigma) = 
\min\left\{1, \frac{2}{\beta_1\log\gamma} 
L_1 \sigma_1^{-\beta_1}\right\}$.
\end{lemma}

\bigskip
The service rate $R=\min\{R_1,R_2/\gamma\}$ in the expression for the 
service curve is the result of a min-plus convolution of the
service curves at the individual nodes.
The logarithmic term can be removed at the expense
of lowering the exponent using Eq.~(\ref{eq:remove-log}) from Lemma~\ref{lemma:K-sigma-alpha}. 
If the second  node also offers an \htt service curve,
with $\eps_2 (\sigma)=L_2\sigma^{-\beta_2}$, then
for every choice of $\beta$ 
with $\beta<\beta_1$ and $\beta\le \beta_2$ 
there exists a constant $L=L(\beta,R_2,\gamma)$ such that
$\eps(\sigma) \le L\sigma^{-\beta}$.
The value of the constant $L$ can be computed from
Lemma~\ref{lemma:K-sigma-alpha} (Eqs.~(\ref{eq:lower-power}) and~(\ref{eq:minimize-sum})).

\bigskip\noindent
{\sc Proof.} We proceed by  inserting the service guarantee 
for $\Aout_1=\Ain_2$ at the first node
into the service guarantee at the second node.
Similar to the backlog and delay bounds, this requires an
estimate for an entire sample path of the service at
the first node. 

Fix $t\ge 0$.  We consider discretized time points $t-y_k$, 
where $y_0=0$ and $y_k=\tau+\gamma' y_{k-1}$
for some $\tau>0$ and $\gamma'>1$ to be chosen below. 
For $t-y_k\le s<t-y_{k-1}$, we have
$$
\Ain_2(s)+\left[R_2(t-s)-\sigma\right]_+
\ge 
\Ain_2\left(t-y_k\right) +\left[R_2y_{k-1}-\sigma\right]_+\,,
$$
and thus
\begin{align}
\Ain_2\conv \S_2(t;\sigma) 
&\geq \inf_{k\geq1} \left\{ \Ain_2\left(t-y_k\right) 
+\left[\frac{R_2}{\gamma'}y_k-
\Bigl(\sigma+\frac{R_2}{\gamma'}\tau\Bigr)\right]_+\right\}.
\label{eq:conv1}
\end{align}
Set $R=\min\left\{R_1,\frac{R_2}{\gamma}\right\}$
and let $\gamma'>1$ and $\delta>0$ be chosen so that
$\frac{R_2}{\gamma'}-\delta=R$. Also fix $\sigma_1,\sigma_2>0$
and set $\sigma=\sigma_1+\sigma_2$.
If for a given sample path
\begin{equation}
\label{eq:conv1b}
\Aout_2(t)\geq \Ain_2\conv \S_2(t;\sigma_2)
\end{equation}
and, for all $k\geq1$ with $y_k\le t$, 
\begin{equation}
\label{eq:conv2}
\Aout_1(t-y_k)\geq
\Ain_1 \conv \S_1 (t-y_k; \sigma_1 + \delta y_k -\frac{R_2}{\gamma'}\tau) \ , 
\end{equation}
then we can insert the lower bound for $\Aout_1=\Ain_2$ from
Eq.~(\ref{eq:conv2}) into Eq.~(\ref{eq:conv1}). 
After collecting terms, the result is
$\Aout_2(t) \geq \Ain_1\conv \S(t;\sigma)$.

The violation probability of Eq.~(\ref{eq:conv1b}) is given by 
$\eps_2(\sigma_2)$. 
Assume for the moment that $\sigma \ge \frac{\delta\tau}{\gamma'-1}$.
We estimate  the violation probability of Eq.~(\ref{eq:conv2}) by
\begin{eqnarray*}
\notag &&\hskip-1cm
\P\bigl(\mbox{Eq.~(\ref{eq:conv2}) fails for some $k$ with $y_k\le t$} 
\bigr)\\
\notag &\le& L_1\sum_{k=1}^\infty
\P\bigl(
\Aout_1(t-y_k)<\Ain_1\conv  \S_1(t-y_k;\sigma_1+\delta y_k
-R_2\tau/\gamma')\bigr) \\
&\le& \frac{L_1}{\log\gamma'}
\left(\frac{1}{\beta_1} +
\left[\log \frac{ (\gamma'-1)(\sigma_1-R\tau)}{\delta\tau} 
\right]_+ \right) (\sigma_1-R\tau)^{-\beta_1} \,.
\end{eqnarray*}
In the first step, we  have used the union bound 
and the \htt\ service curve $\S_1$.
In the second step, we have used Lemma~\ref{lm:est2} to evaluate the
sum (with $\gamma'$ in place of~$\gamma$, and
$a=\delta$), and 
recalled that $R_2/\gamma'-\delta=R$. 
(Here, we have used the assumption on $\sigma$ given before the equation).
We eliminate the shift with Lemma~\ref{lemma:K-sigma-alpha} (Eq.~(\ref{eq:remove-shift})),
and insert the optimal choice 
$\tau= R^{-1} \left(\frac{L_1}{\beta_1\log\gamma'}
\right)^{\frac{1}{\beta_1}}$.
Taking $ \gamma'=\sqrt{\gamma}$ and $\delta=R(\gamma'-1)$,
we arrive at
$$
\P\!\left(
\begin{array}{c}
\mbox{Eq.~(\ref{eq:conv2}) fails for}\\
\mbox{some $k$ with $y_k\le t$}\end{array}
\right) \le
\tilde L_1\sigma_1^{-\beta_1}
\Bigl(\log(\tilde L_1 \sigma_1^{-\beta_1})+2\Bigr)
\le \tilde{\eps}_1(\sigma_1)\bigl(
|\log \tilde{\eps_1}(\sigma_1)|+2\bigr)\,,
$$ 
where $\tilde L_1=\frac{2^{\max\{1,\beta_1\}}}{\beta_1\log\gamma}L_1$.
This bound remains valid for $\sigma< \frac{\delta \tau}{\gamma' - 1 }= R\tau$, 
since then we have $\tilde \eps_1(\sigma)=1$.
Applying the union bound to the violation probabilities 
in Eqs.~(\ref{eq:conv1b}) and~(\ref{eq:conv2}) gives the claim of the lemma.
\hfill$\Box$

\medskip\noindent Iterating the lemma results in the following
end-to-end service guarantee, referred to as {\em network service curve}. 
To keep the statement of the theorem simple, we have
assumed that each node offers an \htt service guarantee with the
same rate $R$, the same constant $L$, and
the same power law $\beta$.  The general case can be reduced
to this with the help of Lemma~\ref{lemma:K-sigma-alpha}
(Eqs.~(\ref{eq:lower-power}) and (\ref{eq:minimize-sum})).

\bigskip
\begin{theorem}{\sc \htt Network Service Curve. }
\label{th:convolution} Consider an arrival flow
traversing $N$ nodes in series, and assume that the service
at each node $n=1,\dots, N$ satisfies an \htt service curve 
\[
\S_n(t,\sigma)=[R t-\sigma]_+\ ,  ~~\eps(\sigma)=L \sigma^{-\beta} \ .
\]
Then, for every choice of $\gamma>1$, the network provides the 
service guarantee
\begin{eqnarray*}
\S_{net} (t,\sigma) & = & \bigl[(R/\gamma)t-\sigma\bigr]_+\, , \\
\eps_{net}(\sigma) 
&\le& N^{2+\beta} \cdot 2^{[\beta-1]_+}
\cdot
\tilde \eps(\sigma)
\left(|\log \tilde \eps(\sigma)| + (1+\beta)\log N
+ 2 \right)
\,,
\end{eqnarray*}
where $\tilde \eps(\sigma) = \min\left\{1, 
\frac{2^{\max\{1,\beta\}}}{\beta 
\log\gamma} L \sigma^{-\beta}\right\}$.  
\end{theorem}

\bigskip 
\noindent{\sc Proof.} We use Lemma~\ref{lm:convolution}
to recursively estimate the service offered by the 
last $n$ nodes with  $n=2,\dots ,N$. In each step, we reduce the
service rate by a factor $\gamma^{\frac{1}{N-1}}$ in place of~$\gamma$.
Fix $\sigma$, and set $\sigma_n=\sigma/N$ for
$n=1,\dots N$. 
If $\tilde \eps(\sigma/N)\ge 1$, there is nothing 
to show. Otherwise, we obtain
\begin{align*}
&\hskip-0.4cm 
\P\bigl( \Aout_N(t) < \Ain_1 \conv \S_{net} (t; \sigma)  \bigr) 
\le \sum_{n=1}^N  N\tilde\eps(\frac{\sigma}{N})
\bigl(|\log(N\tilde\eps(\frac{\sigma}{N})|+2\bigr)\,,
\end{align*}
and the claim follows by collecting the factors of $N$.
\hfill$\Box$

\begin{figure}[t]
\centerline{\includegraphics[width=0.4\textwidth]{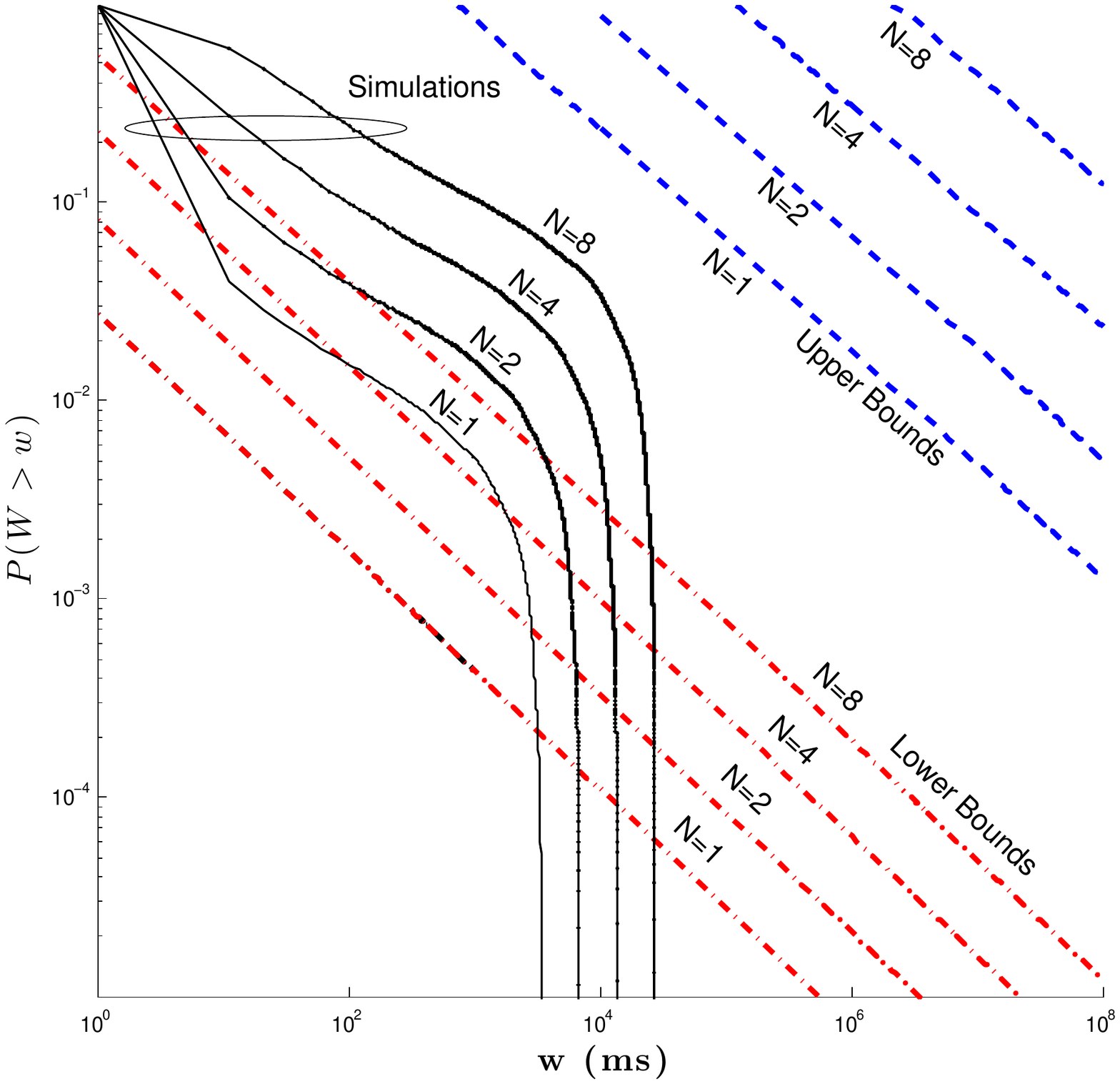}} \vspace{-10pt} 
\caption{Log-log plot of delay bounds for $N$ nodes.} \label{fig:multinodedelay}
\end{figure}

\bigskip
\noindent{\bf Example: }
We perform a multi-node delay analysis for a sequence of homogeneous 
nodes with the same parameters used for Fig.~\ref{fig:loglogdelay}.
The reason for using this example, is that it permits us to draw 
a comparison with 
the lower bound for multi-node networks from \cite{BuLiCi06} given 
in Eq.~(\ref{eq:lower-quant-Pareto}). 
In Fig.~\ref{fig:multinodedelay} we show lower and upper bounds 
for networks with $N=1,2,4,8$ nodes. For reference, we also include the 
results of individual simulation runs with $10^8$ packets. 
The difference between lower and upper bounds is more pronounced than in 
the single-node analysis, and increases with the number of 
nodes $N$. For both lower and upper bounds, the straight lines make the power-law decay in $w$ apparent.
The growth of the bounds in  $N$ suggests a power-law growth in $N$, where 
the larger spacing for the upper bounds indicates a higher power. 
As before, we see that simulations violate analytical lower bounds. 
Since  simulations of heavy-tailed traffic have 
little predictive values for larger delays,  
our analytical bounds provide more reliable  estimates, even with the 
significant gap between upper and lower bounds. 

\section{Scaling of Delay Bounds}\label{sec:scaling} 

We now explore the scaling properties of the delay bounds
from the previous section. 
Throughout this section,
we consider a network as in Fig.~\ref{fig:crossTraffic}.
We assume that the network is homogeneous,
in the sense that all nodes have the same capacity $C$,
and all traffic is bounded by
\htss envelopes as in Eq.~(\ref{eq:envelope})
with the same power $\alpha$ and Hurst parameter $H$.
The cross traffic at each node
has rate $r_c$ and constant $K_c$,
and the through flow has 
rate $r_0$ and constant $K_0$.
Traffic can be either fluid-flow or packetized. In the latter case, 
the packet-size distribution of the through flow
satisfies
$$
\P\bigl\{X>\sigma\bigr)\le L_p\sigma^{-\alpha_p}\,.
$$
We assume the stability condition $r_0+r_c<C$ holds at each node.

\medskip\noindent 
{\bf Single node, large delays ($w\to\infty$). }
Our first result  concerns the power-law decay 
of the delay distribution at a single node.
We choose a relaxation of $\mu =\frac{1}{2}(C-r_c-r_0)$,
and use the leftover service curve
from Eq.~(\ref{eq:leftoverservicegeneral}),
given by
\begin{equation}\label{eq:S-node}
\S(t;\sigma)=[(C-r_c-\mu)t-\sigma]_+\,,\;
\eps_s(\sigma)\le L\sigma^{-\beta}\,,
\end{equation}
where 
\begin{equation}
\label{eq:def-beta}
\beta=\min\{\alpha_p-1,\alpha(1-H)\}\,,
\end{equation}
and $L$ is an explicitly computable constant. (For fluid-flow traffic, that is, without a packetizer, the first term 
does not appear, and we have $\beta = \alpha(1-H)$.) 
We then apply the delay bound of Theorem~\ref{th:ppb}
with $R=r_0+\mu$ to obtain
\begin{equation}
\P(W(t)>w)\le M R^\beta w^{-\beta}\, .
\label{eq:tail-single}
\end{equation}
The constant $M$ is determined by Eq.~(\ref{eq:yy}) 
of Theorem~\ref{th:ppb} with $\beta'=\beta$.  
This shows that the delay decays with the same power law
as the backlog bound in Eq.~(\ref{eq:B-rigorous}).

\medskip\noindent 
{\bf Multiple nodes, large delays ($w\to\infty$). }
Now we consider scaling in networks with 
$N>1$ nodes. We  choose
$\mu=\frac{1}{3}(C-r_c-r_0)$.
We obtain at each node the service curve in Eq.~(\ref{eq:S-node}),
with $\beta$ given by Eq.~(\ref{eq:def-beta}).
We next choose $\gamma=\frac{C-r_c\mu}{r_0+\mu}$
and obtain from
Theorem~\ref{th:convolution} the network service curve
$$
\S_{net}(t;\sigma)=
\bigl[R_{net}t-\sigma\bigr]_+\,,
$$
where $R_{net}= r_0+\mu$,
and with violation probability bounded by
$$
\eps_{net}(\sigma)\le 
N^2 \left([\log z]_+ + \frac{2}{\beta}\right)z^{-\beta}\Bigg\vert_{
z= \frac{\sigma}{\tilde L^{1/\beta}N}}\,,
$$
with an explicitly computable constant $\tilde L$
that does not depend on $N$.
Combining the network service curve with
the arrival envelope, we obtain from Eq.~(\ref{eq:ppb})
of Theorem~\ref{th:ppb} for the end-to-end delay $W_{net}$ that 
$$
\P\bigl(W_{net} (t) >w\bigr)
\le  \inf_{\sigma_1+\sigma_2=R_{net}w}
\left\{\tilde K \sigma_1^{-\beta} + 
\eps_{net}(\sigma_2)\right\}\,.
$$
Here, the constant $\tilde K$ is given by Eq.~(\ref{eq:def-constant})
with  $r_0$ in place of $r$.
We further choose
$\sigma_1=N^{-1-\frac{2}{\beta}} R_{net}w$ and 
$\sigma_2= R_{net}w-\sigma_1$, and see that 
$$
\P(W_{net}(t)>w) \le N^{2+\beta}\left(
M_1 \log w + M_2\log N +M_3\right)
w^{-\beta}\,.
$$
The constants $M_1$, $M_2$, and $M_3$
are again explicitly computable, and
do not depend on $N$.
The tail of the delay distribution, i.e., when $w\to\infty$, 
is dominated by the
first summand in the brackets,
thus, we have the asymptotic upper bound 
\begin{equation}
\label{eq:tail-multi}
\P(W_{net}(t)>w) = 
O\bigl(w^{-\beta}\log w\bigr)\,,\quad (w\to\infty)\,.
\end{equation}

\medskip\noindent 
{\bf Multiple nodes, long paths ($N\to\infty$). }
For long paths, i.e., $N\to\infty$, the second summand dominates.
The quantiles of the delay, defined by
$$
w_{net}(\eps)=\inf\bigl\{ w>0\ \mid 
\P(W_{net}>w)\le \eps\bigr\}
$$
satisfy
\begin{equation}
\label{eq:quantiles}
w_{net}(\eps) = 
O\bigl(N^{\frac{2+\beta}{\beta}}(\log N)^{\frac{1}{\beta}}\bigr)\,,
\quad (N\to\infty)\,.
\end{equation}

\medskip\noindent 
{\bf Comparison of scaling bounds. }
We next compare these upper bounds with scaling 
results from the literature for a Pareto service time distribution
and no cross traffic, where 
traffic arrives in the form of evenly spaced
packets $X_i$, with an i.i.d. Pareto
packet-size distribution, as characterized in 
Section~\ref{sec:envelope}. We assume that  
service times of packets are identical at each node in the sense of \cite{Boxma79}. 
By scaling the units of time and traffic, we may assume an average packet size of $E[X]=1$ and a 
link rate $C=1$, resulting in a rate $\lambda = \rho$, 
where $\rho$ is the utilization. 

For this model, it is known from queueing theory that
the delay at a single node decays with a power law with
exponent $\alpha-1$~\cite{Cohen73}.
Theorem~1 from \cite{Cohen73} yields for the
queueing time of the $k$-th packet in the steady state
$Q=\lim_{k\rightarrow\infty}Q_k$ that 
$$
\P\Big(Q>\sigma\Big)\sim 
\frac{\rho}{1-\rho} 
\frac{(\alpha-1)^{\alpha-1}}{\alpha^\alpha }
\sigma^{-(\alpha-1)} \,,\quad (\sigma\to\infty)\,.
$$
The delay of the $k$th packet is the sum of its queueing time $Q_k$ and
its processing time $X_k$. 
This per-packet delay is related with
the delay $W(t)$ at a given time by
$$
W(t)= \Bigl(Q_{k(t)} + X^*(t)\Bigr)I_{B(t)>0}\,,
$$
where $k(t)$ is the number of the packet  
being processed at time~$t$, and $X^*(t)$ is 
the lifetime of the current packet, as 
defined in Section~\ref{sec:service}. 
Since packets are i.i.d., $W_{k(t)}$ is independent of 
$X^*(t)$ and its distribution agrees with $W_k$,
and we can compute
\begin{equation}
\lim_{t \rightarrow \infty }\P\bigl( W(t)>w \bigr)
\sim \frac{\rho}{1-\rho} c(\alpha)
w^{-(\alpha-1)} \,,\quad (w\to\infty)\,,
\label{eq:exactWt}
\end{equation}
where $c(\alpha)$ is a constant that depends on the 
tail index.

If we compare this asymptotic exact result with our bound from
Eq.~(\ref{eq:tail-single})~and(\ref{eq:tail-multi}), we see that
$\beta=\alpha-1$, and so Eq.~(\ref{eq:tail-multi})
provides (up to a logarithmic correction) the same  power-law decay
as  Eq.~(\ref{eq:exactWt}). The constant $M$ 
in Eq.~(\ref{eq:tail-single}) 
is of order $O\bigl((1-\rho)^{-2}\bigr)$, while the right hand side
of Eq.~(\ref{eq:exactWt}) is of order $(1-\rho)^{-1}$,
which indicates that our delay bound becomes
pessimistic as $\rho\to 1$. 

Exploring the scaling in a tandem network, we first note that 
Eq.~(\ref{eq:def-constant-0}) states that
for a single node, the tail probability of the delays 
decays with $O\bigl(w^{-(\alpha-1)}\log w\bigr)$.
Since end-to-end delays exceed 
the delay at a single node, Eq.~(\ref{eq:exactWt})
guarantees that $W(t)=\Omega\bigl(w^{-(\alpha-1)}\bigr)$.
Thus, our upper and lower bound differ by at most a logarithmic factor.
Eq.~(\ref{eq:quantiles}) implies
furthermore that delay quantiles are bounded by
$O\bigl(N^{\frac{\alpha+1}{\alpha-1}}(\log N)^{\frac{1}{\alpha-1}}
\bigr)$ as $N\to\infty$.
From the lower bound from ~\cite{BuLiCi06} 
given in Eq.~(\ref{eq:lower-quant-Pareto})
we can obtain that quantiles of the 
end-to-end delay grow at least as fast as
$ w_{net}(\eps)=  \Omega\bigl(N^{\frac{\alpha}{\alpha-1}}\bigr)$.  

Lastly, we note that end-to-end delays are expected
to grow more slowly if service
times are independently regenerated 
at each node.  A large buffer asymptotic 
from \cite{BaccelliLF04} for multi-node networks could be used 
to obtain the scaling properties of such a network. 

\section{Conclusions}\label{sec:conclusion}
We have presented an end-to-end analysis of networks with heavy-tailed and 
self-similar traffic. 
Working within the framework  of the network calculus, we 
developed envelopes for heavy-tailed self-similar 
traffic and service curves for heavy-tailed service models. 
By presenting new sample path bounds for arrivals and service, 
we were able to derive 
non-asymptotic performance bounds on backlog and delay, as well as 
network-wide service characterizations. 
We explored the scaling behavior of the derived bounds and 
showed that, for single nodes,  the tail probabilities of 
our delay bounds observe the same power-law decay as known 
results for G/G/1 systems. We also described 
the scaling behavior of end-to-end delays. 
Our paper may motivate further study of the  conditions 
under which  performance bounds in a heavy-tailed regime can be tightened.  
A useful, possibly difficult extension is  
the derivation of a multi-node service curve that accounts for self-similarity, in addition to 
heavy-tails.

\section*{Acknowledgments}
We thank Prof. Anja Feldmann at TU Berlin for giving us 
access to a collection of data trace. J. Liebeherr thanks Deutsche Telekom Laboratories for 
hospitality during a sabbatical. 
We thank Prof. Stephen McLaughlin for sharing a software library 
for alpha-stable distributions. We thank Yashar Ghiassi for comments on the manuscript.

\newcommand{\noopsort}[1]{}

\newpage
\appendix

\vspace{2cm} 

\section {Self-Similar Traffic with Gaussian Tails}

While self-similar traffic is 
expected to be heavy tailed, some self-similar
traffic types appearing in network analysis exhibit a 
faster decay. In the following we present a derivation of a network 
service curve for traffic with a Gaussian bound on the violation probability.  
More precisely, will consider processes that satisfy
for all $\sigma>0$
\begin{equation}
\label{eq:ss-2}
\P \Big(A(s,t) > r(t-s)+\sigma (t-s)^H\Big)  \leq 
K e^{-\frac{1}{2}(\sigma/b)^2}\, , 
\end{equation}
with some constants $K>0$, $a>0$, and $\alpha\ge 0$.
In terms of Eq.~(\ref{eq:localenv1}), this corresponds to 
a statistical envelope
\begin{equation}
\label{eq:localenv2}
\G(t;\sigma)=rt+\sigma t^H\,,\quad \eps(\sigma) = 
K e^{-\frac{1}{2}(\sigma/b)^2}\,.
\end{equation}
To motivate that envelopes given by
Eq.~(\ref{eq:localenv2}) can provide useful traffic models,
consider first the fractional Brownian motion
(fBm) traffic model.  In this model, arrivals are given by 
\begin{equation}\label{eq:asss-2}
\Ain(t)\stackrel{dist.}{=} rt+bt^{H} N_{0,1}~, 
\end{equation}
where $N_{0,1}$ is the standard normal distribution, 
and $b$ is a parameter that relates to the standard deviation. 
Eq.~(\ref{eq:asss-2}) is the special case of the $\alpha$-stable 
process from Eq.~(\ref{eq:asss}) with $\alpha = 2$. 
Using that the density of $N_{0,1}$ is given by
$\phi(x)=\frac{1}{\sqrt{2\pi}}e^{-\frac{1}{2}x^2}$,
we see from Eq.~(\ref{eq:asss-2}) that
\begin{equation}
\P \Big(A(s,t) > r(t-s)+\sigma (t-s)^H\Big)  
= \int_{\sigma}^{\infty} \frac{1}{\sqrt{2\pi}} e^{-\frac{1}{2}x^2}\, dx
\le \frac{1}{2}e^{-\frac{1}{2}\sigma^2}\,,
\label{eq:ss-env-2}
\end{equation}
i.e., Eq.~(\ref{eq:localenv2}) with $K=1/2$
is a statistical  envelope.

An alternative description of arrival processes
with Gaussian tails is through their
{\em effective bandwidth}, defined by
$$
eb(\theta, t)
=\frac{1}{\theta t} \log \E\left(e^{\theta A(t)}\right)\,.
$$
The effective bandwidth of the self-similar 
process in Eq.~(\ref{eq:asss-2}) is given by
$ eb(\theta,t)= \rho + \frac{1}{2} b^2\theta t^{2H-1}$.
Let us consider more generally processes whose 
effective bandwidth satisfies the self-similar bound
\begin{equation}
\label{eq:effBW}
eb(\theta, t)
\le \rho + \frac{1}{2} b^2\theta t^{2H-1}\,.
\end{equation}
By the Chernoff bound, Eq.~(\ref{eq:effBW}) implies
that
$$
\P \Big(A(s,t) > r(t-s)+\sigma (t-s)^H\Big)  \leq 
e^{-\frac{1}{2}(\sigma/b)^2}\,,
$$
i.e., Eq.~(\ref{eq:localenv2})
with $K=1$ is a statistical envelope.
Note that for $H>\frac{1}{2}$ and any given choice of $\theta$,
Eq.~(\ref{eq:effBW}) does not define a linear bounded envelope
process as defined in~\cite{Chang94}.

\bigskip 
\begin{lemma} {\sc sample path envelope for self-similar processes
with Gaussian tail. } 
\label{lm:arrival-samplepath-Gaussian}
Assume that arrivals to a flow are bounded by a statistical
envelope, given by
\begin{equation}
\G(t;\sigma)=rt+\sigma t^H \ , ~~ 
\eps(\sigma) = K e^{-\frac{1}{2}(\sigma/b)^2} \ . 
\end{equation}
Then, for any choice of
$\mu>0$, a statistical sample path envelope for the arrival 
process is given by
$$
\overline{\G}(t;\sigma)=(r+\mu)t+\sigma\,,\quad
\bar \eps(\sigma) =
L e^{-(\sigma/c)^\beta}\,,
$$
with parameters
$\beta=2(1-H)$, $c = \left(\frac{2b}{\mu^H}\right)^{\frac{1}{1-H}}$,
and $L= e\cdot\max\left\{1, 4^HK\frac{r/\mu+2-H}{H(1-H)}\right\}$.
\end{lemma}

\bigskip
{\sc Proof.} Fix $\mu>0$, let $1<\gamma<1+\frac{\mu}{r}$,
and set $b=\frac{r+\mu}{\gamma}-r>0$.
As done for the backlog bound from Eq.~(\ref{eq:B-rigorous}), we argue that  
\begin{eqnarray}
\notag
\P \Big(\sup_{s \leq t } \Big\{A(s,t) - \overline{\G}(t - s; \sigma)\Big\} 
> 0 \Big) &\le& 
\sum_{k=-\infty}^\infty
\eps\left(\frac{\sigma+ bx_k}{x_k^H}\right)\\
&\le& \frac{1}{H(1-H)\log\gamma}
\int_z^\infty \frac{\eps(x)}{x}\, dx \\
\notag
&\le& 
\frac{K}{H(1-H)\log \gamma} z^{-2}e^{-\frac{1}{2}z^2} \,,
\label{eq:Weibull-3}
\end{eqnarray}
where $z= \frac{b^H \sigma^{1-H}}{\gamma^{H(1-H)}}$.
Here, the first inequality follows by discretization.
In the second step,  we have
used Lemma~\ref{lm:est1} 
and replaced the integration variable $x$ with $x/b$.
In the third step, we have evaluated the integral.

We want to replace the variable $z$ by
$y=\mu^H\sigma^{1-H}/b$,
so that $\frac{1}{2}y^2=(\sigma/c)^\beta$.
Suppose for the moment that
$y^2\ge 2$, and choose $\gamma$ such that
$\log\gamma= (r/\mu +2-H)^{-1}y^{-2}$.
Then $\gamma<1+r/\mu$, as required. 
Moreover,
$$z^2 = y^2\left(\frac{b}{\mu\gamma^{1-H}}\right)^{2H}
\ge y^2\left(1-(r/\mu +2-H)\log\gamma\right)^{2H} \ge \frac{1}{2}y^2\,,
$$
and 
$$
z^2 \ge  y^2\left(1-2(r/\mu +2-H)\log\gamma\right)\\
\ge  y^2-2\,,
$$
and we obtain the bound
$$
\P \Big(\sup_{s \leq t } \Big\{A(s,t) - \overline{\G}(t - s; \sigma)\Big\} 
> 0 \Big)
\le Le^{-(\sigma/c)^\beta}\,.
$$
This proves the claim for $y^2\ge 2$.
On the other hand, for $y^2<2$, the claim holds trivially
since the right hand side exceeds $1$.
\hfill $\Box$

\bigskip 
As in Section~\ref {subsec-leftover}, the sample path envelope immediately
yields a leftover service curve.  If fluid-flow traffic 
arrives to a link of constant rate $C$,
where it is subject to cross traffic that has a
statistical envelope with Gaussian tails, as in
Eq.~(\ref{eq:localenv2}), then
for every choice of $\mu>0$,
\begin{equation}
\label{eq:servcurve-weibull}
\S(t;\sigma)=[(C-r_c-\mu)t-\sigma]_+\,,\quad 
\eps(\sigma) = L e^{-(y/c)^\beta}
\end{equation}
provides a statistical service curve for the through flow with a Weibullian bound on the 
violation probability. Here, the parameters $\beta$, $c$, and $L$ are 
given by  Lemma~\ref{lm:arrival-samplepath-Gaussian}.
Next, we show how to concatenate
such service curves:

\bigskip 
\begin{lemma}{\sc Concatenation of Weibull-tailed service curves. }  
\label{lm:convolution-Weibull} 
Consider two consecutive nodes on the path of a flow through a network.
Assume that the first node offers a statistical service curve
$\S_1(t;\sigma)= [R_1t-\sigma]_+$ with $\eps_1(\sigma)= L_1e^{-(\sigma/c_1)^{\beta_1}}$,
and the second node offers a statistical service curve
$\S_2(t;\sigma)= [R_2t-\sigma]_+$
with an arbitrary function $\eps_2(\sigma)$. Then, for every choice of
$\gamma>1$, the two nodes offer the 
combined service curve given by
\begin{eqnarray*}
\S(t;\sigma) &=& \bigl[\min\{R_1,R_2/\gamma\}-\sigma\bigr]_+\,,\\
\eps(\sigma) &=& \inf_{\sigma_1+\sigma_2=\sigma}
\left\{ 
\tilde L_1 e^{-(\sigma_1/c_1)^{\beta_1}} + \eps_2(\sigma_2) \right\}
\,,
\end{eqnarray*}
where $\tilde L_1 = \max\left\{e^2, 
\frac{\gamma}{\gamma-1}
\frac{(2e)^{[\beta_1-1]_+}}
{c_1}L_1\right\}$.
\end{lemma}

{\sc Proof.} We follow the proof of Lemma~\ref{lm:convolution},
with the Weibullian function 
$\eps_1(\sigma)= L_1e^{-(\sigma/c_1)^{\beta_1}}$
in place of the power law.  Set $R=\min\{R_1,R_2/\gamma\}$,
$\delta= R_2-R$, and consider the event
\begin{equation}
\label{eq:conv2-Weibull}
\Aout_1(t-y_k)\geq
\Ain_1 \conv \S_1 (t-y_k; \sigma_1 + \delta y_k -R_2\tau) \,,
\end{equation}
where $y_k$ is a sequence of discretization time steps.
We argue as in the proof of Lemma~\ref{lm:convolution} that 
$\S$ is a service curve, with violation probability given by 
$$
\eps(\sigma)\le \inf_{\sigma_1+\sigma_2=\sigma}
\left\{
\P\bigl(\mbox{Eq.~(\ref{eq:conv2-Weibull}) 
fails for some $k$ with $y_k\le t$} \bigr)
+\eps_2(\sigma_2)\right\}\,.
$$
We could now use Lemma~\ref{eq:est2} to bound the violation
probability of Eq.~(\ref{eq:conv2-Weibull}).
However, since $\eps_1(\sigma_1)$ is an integrable function,
we can obtain a slightly stronger bound
by using the arithmetic sequence
$y_k=k\tau$, as in~\cite{CiBuLi06}. The resulting estimate is
\begin{eqnarray*}
\notag \P\bigl(\mbox{Eq.~(\ref{eq:conv2-Weibull}) 
fails for some $k$ with $y_k\le t$} \bigr)
\notag &\le& \sum_{k=1}^\infty
\eps_1\bigl(\sigma_1 +\delta y_k -R_2\tau \bigr) \\
\notag &\le& \frac{1}{\delta\tau}\int_{\sigma_1-R_2\tau}^\infty 
\eps_1\bigl(x)\, dx \\
\notag &\le& L_1 \frac{c_1}{\beta_1\delta\tau}
z^{-(\beta_1-1)}
e^{-z^{\beta_1}}\Bigg\vert_{z=\frac{\sigma_1-R_2\tau}{c_1}}
\end{eqnarray*}
We have first used the union bound
and the \htt\ service curve, then
replaced the sum by an integral, and
finally evaluated the integral. 
Suppose for the moment that $(\sigma_1/c_1)^{\beta_1}\ge 2 $. We 
choose $\tau=\frac{c_1^{\beta_1}}{\beta_1R_2\sigma_1^{\beta_1-1}}$
and use that
$$
(\sigma_1-R_2\tau)^{\beta_1}\ge 
\sigma_1^{\beta_1}-\max\{1,\beta\}R_2\tau\sigma_1^{\beta_1-1}\,,
\quad \frac{\sigma_1}{\sigma_1-\tau}\le 2
$$
to obtain
$$
\P\bigl(\mbox{Eq.~(\ref{eq:conv2-Weibull}) 
fails for some $k$ with $y_k\le t$} \bigr)
\le \tilde L_1 e^{-(\sigma_1/c_1)^{\beta_1}}\,.
$$
But this inequality clearly also holds for $(\sigma_1/c_1)^{\beta_1}<2 $. 
The lemma follows by combining this with the service guarantee
at the second node.
\hfill$\Box$

\newpage
With this result, we finally present a network service curve that concatenates 
an arbitrary number of Weibullian service curves.  

\bigskip
\begin{theorem}{\sc Weibullian Network Service Curve. }
\label{th:convolution-Weibull} Consider an arrival flow
traversing $N$ nodes in series, and assume that the service
at each node $n=1,\dots, N$ satisfies 
a service curve 
\[
\S_n(t,\sigma)=[R t-\sigma]_+\ ,  ~~\eps(\sigma)=L e^{-(\sigma/c)^\beta} \  .
\]
Then, for every choice of $\gamma>1$, the network provides the 
service curve  
\begin{eqnarray*}
\S_{net} (t,\sigma) & = & \bigl[(R/\gamma)t-\sigma\bigr]_+\, , \\
\eps_{net}(\sigma) 
& \le & N\left(1+\frac{N}{\log\gamma}\right)\tilde L e^{-(\sigma/Nc)^\beta}\,,
\end{eqnarray*}
where $\tilde L = \max\left\{
\frac{e^2}{N}\log\gamma,
\frac{(2e)^{[\beta-1]_+}}{c} L\right\}$.
\end{theorem}

\bigskip 
\noindent{\sc Proof.} See the proof of Theorem~\ref{th:convolution}.
We use Lemma~\ref{lm:convolution-Weibull}
to iteratively derive the service guarantee for the last $n$ nodes.
In each step, we reduce the rate by a factor of $\gamma^{\frac{1}{N-1}}$.
To simplify the formula, we have replaced
$N-1$ with $N$ whenever it appeared convenient, and used that
$\frac{N}{\log\gamma} \le
\frac{\gamma^{\frac{1}{N}}}{\gamma^{\frac{1}{N}}-1}
\le 1+ \frac{N}{\log\gamma}$.
\hfill$\Box$

\bigskip
Consider now a network as in Figure~\ref{fig:crossTraffic} where all flows have envelopes 
with a Gaussian tail, given by Eq.~(\ref{eq:ss-2}). Assume that the 
network is homogeneous, i.e., all nodes have the same link rate, and 
the cross traffic has the same parameters at each node. 
Combining the results from this section, we obtain, similarly to the analysis 
in Section~\ref{sec:scaling}, the end-to-end delay bound  
$$
\P(W_{net}(t)>w) = 
O\bigl(N^2 e^{-(\frac{w}{Nw_0})^\beta} \bigr)\,,\quad (w\to\infty)\,,
$$
where $M$ and $w_0$ are constants that do not depend on $N$, and $\beta=2(1-H)$. 
It follows that, over long paths, the quantiles of the delays scale as 
$$
w_{net}(\eps) = O\bigl(N (\log N)^{1/\beta}\bigr)\,,
\quad (N\to\infty)\,.
$$
For $\beta = 1$, this bounds recovers the $O (N \log N)$ bound for exponential 
traffic from \cite{CiBuLi06}.
\clearpage

\section{Technical Lemmas}

In our derivations, we frequently use 
properties of the function $\eps(\sigma)=K\sigma^{-\alpha}$ that
appears in the definition of the
\htss envelope. The properties are summarized in the following lemma, 
and presented without proof.
\bigskip

\noindent
\begin{lemma}  
\label{lemma:K-sigma-alpha}
\begin{enumerate}
\item[]
\item {\em (Lower power.) }
We can lower the power
by using that for $K\sigma^{-\alpha}\le 1$ and
$\alpha'<\alpha$
\begin{equation}
\label{eq:lower-power}
K\sigma^{-\alpha}\le 
K^{\frac{\alpha'}{\alpha}} \sigma^{-\alpha'}\,.
\end{equation}

\item {\em (Eliminate logarithmic factor.) }
For $\beta'< \beta$, 
\begin{equation}
\label{eq:remove-log}
\sigma^{-\beta}\log\sigma
\le \frac{1}{e(\beta-\beta')}\sigma^{-\beta'}\,.
\end{equation}

\item {\em (Remove shift.) }
For $\alpha>0$, $\sigma_0>0$, and  $K(\sigma-\sigma_0)^{-\alpha}\le 1$,
we can remove a negative shift by
\begin{equation}\label{eq:remove-shift}
K(\sigma-\sigma_0)^{-\alpha}
\le 2^{[\alpha-1]_+}
\bigl(K+\sigma_0^\alpha\bigr)\sigma^{-\alpha}\,.
\end{equation}

\item {\em (Minimize sum.) }
We can minimize sums of such functions by
\begin{eqnarray}
\min_{\sigma_1+\dots+\sigma_n=\sigma}
\sum_{j=1}^n K_j \sigma_j^{-\alpha} 
&=& \Bigl(\sum_{j=1}^n K_j^{\frac{1}{1+\alpha}}\Bigr)^{1+\alpha}
\sigma^{-\alpha} 
 \le n^\alpha \overline{K}\sigma^{-\alpha}\,,
\label{eq:minimize-sum}
\end{eqnarray}
where $\overline{K} = \frac{1}{n}(K_1 + K_2 + \ldots + K_n)$.
\end{enumerate}
\end{lemma} 

\bigskip
\bigskip

The following derives auxiliary estimates  for two sums that
involve geometric sequences.

\bigskip
\begin{lemma}\label{lm:est1}
Assume that $\eps(x)$ is a nonincreasing nonnegative
function. Fix $\gamma>1$ and $\tau>0$,
and set $ x_k = \tau \gamma^k$.
Then, for every $\sigma\ge 0$ and every $c>0$,, 
\[ %
\sum_{k=-\infty}^{\infty}
\eps\left(\frac{\sigma+ cx_k} {x_k^{H}}\right) 
\leq
\frac{1}{H(1-H) \log\gamma} 
\int_z^\infty \frac{\eps(x)}{x}\, dx \ 
\Bigg\vert_{z=\frac{c^H\sigma^{1-H}}{
\gamma^{H(1-H)}}}\,.
\] %
\end{lemma}
\hfill $\Box$

\bigskip\noindent {\sc Proof.} Consider first the case
where $c=\tau=1$, i.e., $x_k=\gamma^k$.
Each summand in the series satisfies
$$
\eps \left(\frac{\sigma+\gamma^k}{\gamma^{Hk}}\right)
\le\min\bigl\{
\eps\bigl(\sigma\gamma^{-Hk}\bigr), 
\eps\bigl(\gamma^{(1-H)k}\bigr)\bigr\}\,.
$$
Since the first term on the right hand side increases with 
$k$ while the second term decreases, we can bound
the series by the sum of two integrals
\begin{eqnarray*}
\sum_{k=-\infty}^\infty\eps \left(\frac{\sigma+ \gamma^k}
{\gamma^{Hk}}\right)
&\leq& \int_{-\infty}^{T+1} \eps\bigl(\sigma\gamma^{-Ht}\bigr)\, dt
+ \int_T^\infty \eps\bigl(\gamma^{(1-H)t}\bigr) \, dt\,,
\end{eqnarray*}
where the overlap between the intervals of integration 
compensates for the change of monotonicity.
The optimal choice for the limit of integration is
$T= -H + \frac{\log\sigma }{\log\gamma}$, 
so that $\sigma \gamma^{-H(T+1)}= \gamma^{(1-H)t}$. 
In the first integral, the change of variables
$x=\sigma\gamma^{-Ht}$ yields
$$
\int_{-\infty}^{T+1} \eps\bigl(\sigma\gamma^{-Ht}\bigr)\, dt
= \frac{1}{H\log\gamma} \int_z^\infty\frac{\eps(x)}{x}\, dx\,,
$$
where $z= \sigma^{1-H}\gamma^{-H(1-H)}$. 
In the second integral, the change of variables
$x=\gamma^{(1-H)t}$ yields
$$
\int_{T}^{\infty} \eps\bigl(\gamma^{(1-H)t}\bigr)\, dt
= \frac{1}{(1-H)\log\gamma} \int_z^\infty\frac{\eps(x)}{x}\, dx\,,
$$ 
Adding the two integrals proves the claim for $a=\tau=1$.
For other values of $c$ and $\tau$, we rescale
$\sigma = c \tau^{1-H}\tilde \sigma$,
and apply the first case to the function
$\tilde \eps(x)=\eps\bigl (c\tau^{1-H}x\bigr)$.
\hfill $\Box$

\bigskip 

\begin{lemma}\label{lm:est2} Assume that $\eps(x)$ is a 
nonincreasing nonnegative function.
Fix $\tau>0$ and $\gamma>1$, and
define recursively $y_0=0$, $y_k=\tau + \gamma y_{k-1}$. 
Then, for every 
$c>0$ and $\sigma\ge \frac{c\tau}{\gamma-1}$,
$$
\sum_{k=1}^{\infty}\eps \left(\sigma+ c y_k\right)
\leq
\frac{1}{\log\gamma}
\left(\eps(z)\log\Bigl(\frac{\gamma-1}{a\tau}z\Bigr)+ 
+ \int_z^\infty \frac{\eps(x)}{x}\, dx\right)
\Bigg\vert_{z=\sigma+c\tau} 
\,.  \label{eq:est2}
$$
\end{lemma}

\bigskip\noindent  {\sc Proof.} Consider first the case where $c=1$ and 
$\tau=\gamma-1$, i.e.,
$y_k=\gamma^k-1$, and set $z=\sigma+\gamma-1$. For
$\sigma\ge 1$, each summand is bounded by
$$
\eps\left(\sigma+\gamma^k-1\right) 
\le \min \left\{\eps(\sigma+\gamma-1), \eps\left(\gamma^k\right)\right\}\,.
$$
Since  both terms are nonincreasing,
we can bound the series by 
$$
\sum_{k=1}^\infty \eps\bigl(\sigma+\gamma^k-1\bigr)
\le \int_0^T \eps(\sigma+\gamma-1) \, dt
+\int_T^\infty \eps\bigl(\gamma^t\bigr)\, dt\,.
$$
We choose $T=\frac{\log(\sigma+\gamma-1)}{\log \gamma}\ge 1$, 
so that $\gamma^T=\sigma+\gamma-1$, 
and change variables $x=\gamma^t$ in the second
integral to obtain
$$
\sum_{k=1}^\infty \eps\bigl(\sigma+\gamma^k-1\bigr)
\le \frac{1}{\log\gamma}
\left(\eps(z) \log z + 
\int_z^\infty \frac{\eps(x)}{x}\, dx\right)
\Bigg\vert_{z=\sigma+\gamma-1}
\,.
$$
This proves the claim in the special case $a=1$, $\tau=\gamma-1$.
For other values of $c$ and $\tau$, we rescale
$\sigma=\frac{c\tau}{\gamma-1}\tilde \sigma$,
$z =\frac{c\tau}{\gamma-1}\tilde z$,
and apply the first case to
$\tilde \eps(x)=\eps\left(\frac{c\tau}{\gamma-1}x\right)$.
\hfill $\Box$

\end{document}